\pdfoutput=1
\documentclass[cernpreprint,atlasdraft=false,texlive=2016,UKenglish,texmf]{atlasdoc}
 
\usepackage[subfigure]{atlaspackage}
\usepackage{atlasbiblatex}
\usepackage{atlasphysics}
 
\graphicspath{{logos/}{figures/}}
 
\usepackage{multirow}
 
\AtlasTitle{Search for Higgs boson production in association with a high-energy photon via vector-boson fusion with decay into bottom quark pairs at \(\sqrt{s}\) = 13~TeV with the ATLAS detector}
 
\AtlasAbstract{
A search is presented for the production of the Standard Model Higgs boson in association with a high-energy photon.
With a focus on the vector-boson fusion process and the dominant Higgs boson decay into \(b\)-quark pairs, the search benefits from a large reduction of multijet background compared to more inclusive searches.
Results are reported from the analysis of \SI{132}{fb^{-1}} of \(pp\) collision data at \(\sqrt{s}=\SI{13}{TeV}\) collected with the ATLAS detector at the LHC.
The measured Higgs boson signal yield in this final-state signature is \(1.3 \pm 1.0\) times the Standard Model prediction. The observed significance of the Higgs boson signal above the background is \(1.3\) standard deviations, compared to an expected significance of \(1.0\) standard deviations.
}
 
\author{The ATLAS Collaboration}
 
\AtlasRefCode{HIGG-2020-14}
 
\PreprintIdNumber{CERN-EP-2020-179}

\AtlasJournalRef{JHEP 03 (2021) 268}
\AtlasDOI{10.1007/JHEP03(2021)268}
\hypersetup{pdftitle={Search for Higgs boson production in association with a high-energy photon via vector boson fusion at 13 TeV},pdfauthor={The ATLAS Collaboration}}
 
\begin{document}
 
\maketitle
 
\section{Introduction}
Following the discovery of a new particle with a mass of approximately \SI{125}{\GeV} by the ATLAS and CMS Collaborations
at the Large Hadron Collider (LHC)\cite{HIGG-2012-27, CMS-HIG-12-028}, there has been an extensive effort to measure its properties and compare them with theoretical predictions for the Standard Model (SM) Higgs boson.
 
Among the Higgs boson production modes still to be explored, the \(H\gamma\) mode has received attention for its interesting phenomenology~\cite{Gabrielli:2007wf, Gabrielli:2016mdd}, in which the addition of the high-energy photon reorders the usual hierarchy of Higgs boson production cross sections.
Experimental measurements are needed to study the SM predictions and to probe for signs of new physics beyond the SM that affect this production mode, which also has potential as a clean signature for unexpected or invisible Higgs boson decays.
 
The exclusive final state \(H\gamma\) is forbidden in gluon--gluon fusion (ggF) by Furry's theorem~\cite{Furry:1937zz}, but the \(H\gamma\) system may be produced in recoil against a jet.
Contributions from associated \(WH\) and \(ZH\) production are similar in size to the higher-order contributions from ggF.
In vector-boson fusion (VBF), the photon may be radiated either from a charged weak boson or from a scattering quark that showers into a jet (\(j\)).
VBF is the dominant production mode for \(H\gamma\), as its cross section is twice as large as that of all other modes combined.
 
To select the VBF production mode in particular, a search has been developed in the \(H\gamma jj\) final state, with an additional phase-space cut on the jet--jet invariant mass.
To maximize the experimental sensitivity to this final state, the search focuses on the dominant \(H\rightarrow \bbbar\) decay.
A previous VBF \(H(\rightarrow \bbbar)\gamma\) search by the ATLAS Collaboration with \SI{31}{\ifb} of \SI{13}{\TeV} data reported a signal strength measurement, defined as the production rate relative to the SM cross section, of \(\mu_\text{VBF} = 2.5 \pm 1.9 \) and an observed (expected) Higgs boson signal significance of \(1.4\sigma\) \((0.6\sigma)\)~\cite{HIGG-2016-30}.
A complementary VBF \(H(\rightarrow \bbbar)\) search by the ATLAS Collaboration in an all-hadronic signature, without the associated photon, reported an observed (expected) Higgs boson signal significance of \(1.4\sigma\) (\(0.4\sigma\)) in the same dataset~\cite{HIGG-2016-30}.
 
In the time since that search was reported, extensive studies of Higgs production without an associated photon by the ATLAS and CMS Collaborations have now accumulated strong evidence for the VBF production mode and for the \(H\rightarrow \bbbar\) decay.
The ATLAS Collaboration has combined several measurements made with up to \SI{79.8}{\ifb} of \SI{13}{\TeV} data to yield \( \mu_\text{VBF} = 1.21 \pm 0.23\), compatible with the combined CMS measurement of \( \mu_\text{VBF} = 0.73 \pm 0.28\) made with up to \SI{35.9}{\ifb}~\cite{HIGG-2018-57, CMS-HIG-17-031}.
Both the ATLAS and CMS Collaborations have observed the \(H \rightarrow \bbbar\) decay mode, with significances of \(6.7\sigma\) and \(5.6\sigma\) respectively, in decay signatures dominated by contributions from \(WH\) and \(ZH\) production~\cite{Aad:2020jym, CMS-HIG-18-016}.
 
\begin{figure}[htbp]
\centering
\includegraphics[width=0.4\textwidth]{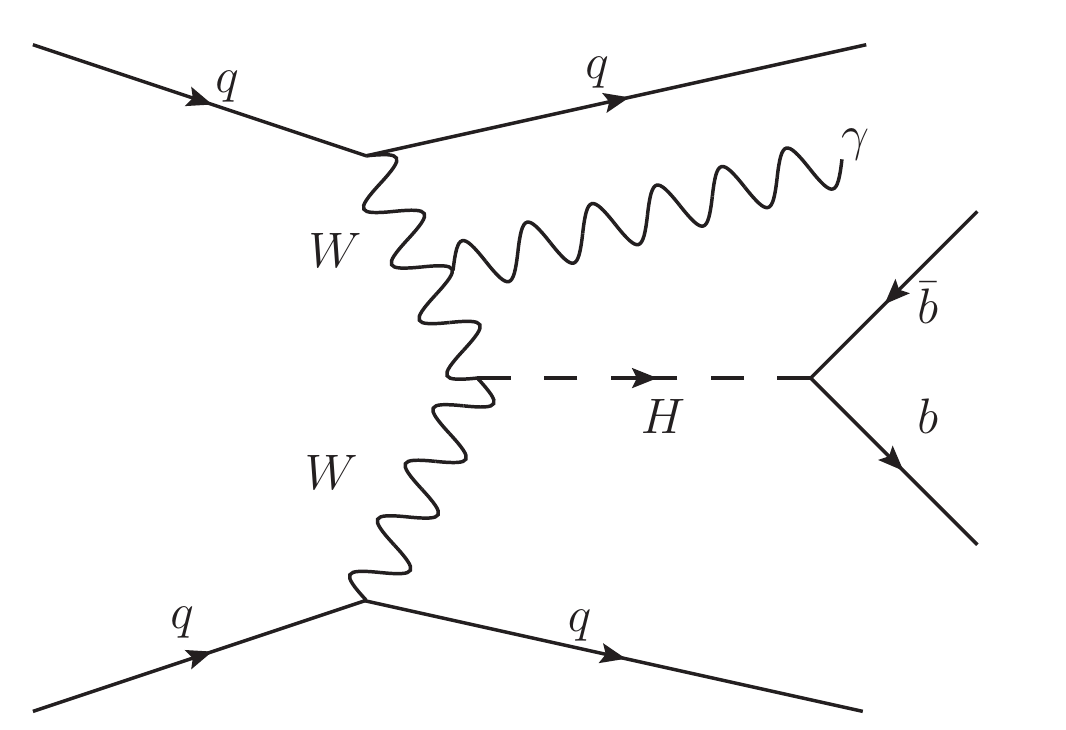}
\includegraphics[width=0.35\textwidth]{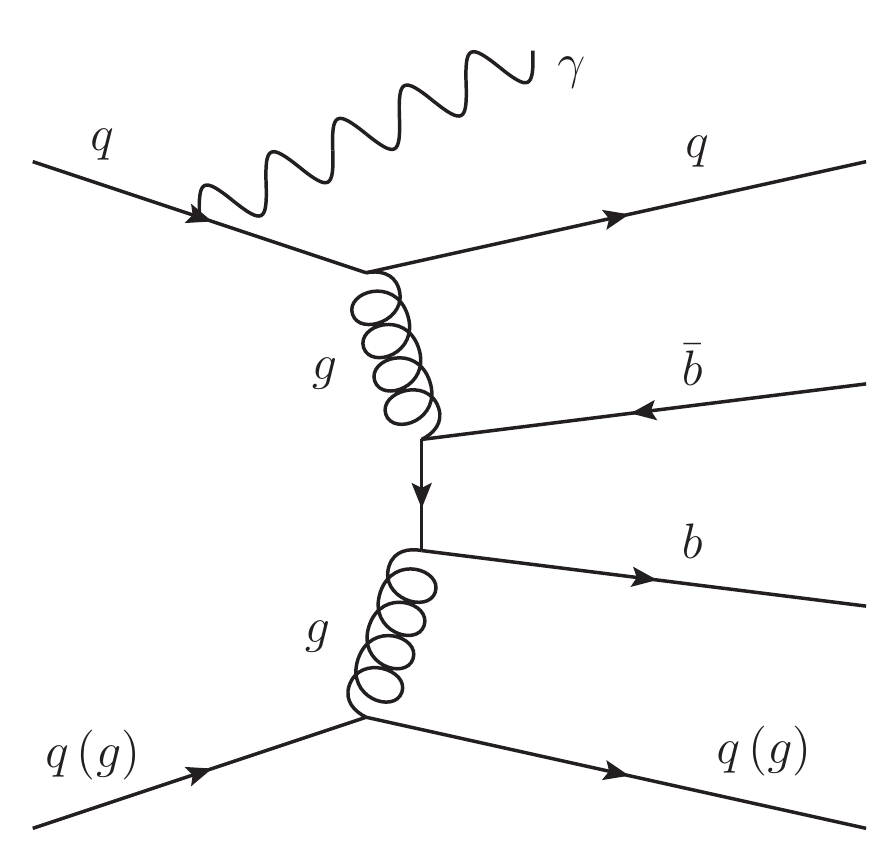}
\caption{Representative leading-order Feynman diagrams for Higgs boson production via vector-boson fusion in association with a photon (left) and the dominant non-resonant \(\bbbar \gamma jj\) background (right).}
\label{fig:diagrams}
\end{figure}

Nevertheless, there are several interesting effects that motivate testing the SM predictions with a measurement of \(H\gamma jj\) production.
First, a unique feature of this production mode is that contributions from \Zboson boson fusion are suppressed, due to destructive interference between matrix elements for initial- and final-state radiation.
This feature allows a more direct measurement of Higgs boson production through \Wboson boson fusion.
Second, the same interference also suppresses the dominant multijet non-resonant \(\bbbar \gamma jj\) background shown in Figure~\ref{fig:diagrams}, improving the expected sensitivity to Higgs boson production.
Third, the presence of an isolated high-energy photon can be used to define efficient trigger algorithms for this search.
 
This paper reports the result of a search for \(H \gamma\) production and subsequent \(H\rightarrow \bbbar\) decay, with a focus on the phase space dominated by the VBF process, using the full Run 2 dataset.
In this phase space, the main backgrounds are non-resonant \(\bbbar \gamma jj\) production and resonant \(Z(\rightarrow \bbbar)\gamma jj\) production.
Kinematic event properties are used in a boosted decision tree (BDT) to classify signal and background events, and the BDT output defines a series of increasingly signal-rich event categories.
The Higgs boson production rate, defined relative to the SM prediction, is extracted from a simultaneous fit to the Higgs boson candidate mass distribution in multiple event categories.
\section{ATLAS detector}
\newcommand{\AtlasCoordFootnote}{
ATLAS uses a right-handed coordinate system with its origin at the nominal interaction point (IP) in the centre of the detector and the \(z\)-axis along the beam pipe.
The \(x\)-axis points from the IP to the centre of the LHC ring, and the \(y\)-axis points upwards.
Cylindrical coordinates \((r,\phi)\) are used in the transverse plane, with \(\phi\) being the azimuthal angle around the \(z\)-axis.
The pseudorapidity is defined in terms of the polar angle \(\theta\) as \(\eta = -\ln \tan(\theta/2)\).
The rapidity is defined as \(y = 0.5 \ln[(E+p_z)/(E-p_z)]\), where \(E\) denotes energy and \(p_z\) is the component of the momentum along the beam direction.
Angular distance is measured in units of \(\Delta R \equiv \sqrt{(\Delta\eta)^{2} + (\Delta\phi)^{2}}\).}
 
The ATLAS experiment~\cite{PERF-2007-01} at the LHC is a multipurpose particle detector with a cylindrical geometry that covers nearly the entire solid angle around the collision point.\footnote{\AtlasCoordFootnote}
It consists of an inner tracking detector surrounded by a thin superconducting solenoid, electromagnetic and hadronic calorimeters, and a muon spectrometer incorporating three large superconducting toroidal magnets.
 
The inner tracking detector covers the pseudorapidity range \(|\eta| < 2.5\).
It consists of silicon pixel, silicon microstrip, and transition radiation tracking detectors immersed in a \SI{2}{\tesla} axial magnetic field.
One significant upgrade for the \(\sqrt{s}=\SI{13}{\TeV}\) run is the insertable B-layer~\cite{ATLAS-TDR-2010-19, PIX-2018-001}, an additional pixel layer close to the interaction point.
 
The calorimeter system covers the pseudorapidity range \(|\eta| < 4.9\).
Within the region \(|\eta|< 3.2\), electromagnetic calorimetry is provided by barrel and
endcap high-granularity lead/liquid-argon (LAr) calorimeters, with an additional thin LAr presampler covering \(|\eta| < 1.8\) to correct for energy loss in material upstream of the calorimeters.
Hadronic calorimetry is provided by the steel/scintillator-tile calorimeter, segmented into three barrel structures within \(|\eta| < 1.7\), and two copper/LAr hadronic endcap calorimeters.
The solid angle coverage is completed with forward copper/LAr and tungsten/LAr calorimeter modules optimized for electromagnetic and hadronic measurements, respectively.
 
The muon spectrometer consists of fast detectors for triggering and high-precision chambers for tracking in a magnetic field generated by superconducting air-core toroids.
The field integral of the toroids ranges between \num{2.0} and \SI{6.0}{\tesla\metre} across most of the detector.
 
A two-level trigger system selects events for offline analysis.
Events that match predefined trigger signatures are selected by the first-level trigger system implemented in custom hardware.
A subset of those events are selected by software algorithms implemented in the high-level trigger~\cite{TRIG-2016-01}.
The first-level trigger accepts events from the \SI{40}{\MHz} bunch crossings at a rate below \SI{100}{\kHz}, and the high-level trigger reduces the rate in order to record events to disk at about \SI{1}{\kHz}.
\section{Data and simulated events}
This search uses LHC proton--proton collision data at a centre-of-mass energy of \SI{13}{\TeV} collected with the ATLAS detector from 2015 to 2018.
This full Run 2 dataset, with the analysis-specific triggers available, corresponds to an integrated luminosity of \SI{132}{\ifb}.
 
Simulated events are used for signal modelling, BDT training, and determining the functional form used to describe the background shape.
The \(H\gamma jj\) signal events were generated at next-to-leading order (NLO) in \alphas with \MGMCatNLO v2.6.2~\cite{Alwall:2014hca}, using the PDF4LHC15 parton distribution function (PDF) set~\cite{Butterworth:2015oua}, and passed to \HERWIG7.1 for parton showering and hadronization using parameter values from the \HERWIG default tunes~\cite{Bahr:2008pv,Bellm:2015jjp}.
This sample of events can effectively be regarded as VBF \(H\gamma\) signal because the contribution from \(VH\) production is less than 0.5\% of the total after event selection.
Higgs boson production from ggF, with Higgs decay into \bbbar, was simulated at next-to-next-to-leading order with \POWHEGBOXV{v2}~\cite{Nason:2004rx,Frixione:2007vw,Alioli:2010xd}, using the CT10 PDF set~\cite{Lai:2010vv} and \PYTHIA~8.212~\cite{Sjostrand:2014zea} for parton showering and fragmentation with the AZNLO tuned parameter set~\cite{STDM-2012-23}.
Contributions from the \(\ttbar H\) process are estimated from \POWHEGBOXV{v2} simulation interfaced with \PYTHIA~8.230.
 
Background events containing two \bjets from the decay of a \Zboson boson, a photon, and two additional jets were generated at leading order (LO) with \MGMCatNLO v.2.3.3 and \PYTHIA~8.210, in separate samples for strong (QCD) and electroweak (EWK) processes.
The dominant source of background events is non-resonant QCD production of at least two \bjets, two other jets, and a photon.
Even though the contribution from these events is modelled with a functional form, a sample of simulated events is used to train the BDT.
This training sample was produced by generating the \(\bbbar \gamma jj\) final state, excluding diagrams containing on-shell \(H\) or \Zboson bosons, at LO with \MGMCatNLO v2.3.3 using the PDF4LHC15 set of PDFs and interfaced to \PYTHIA~8.210 for parton showering and hadronization.
The A14 set of tuned parameters was used for the underlying-event description with the NNPDF2.3LO PDF set~\cite{Ball:2012cx}.
A large sample of non-resonant \(\bbbar \gamma jj\) background events was also produced without parton showering for the background modelling studies described in Section~\ref{sec:modelling}.
For this sample, a parameterized jet energy response function is used to smear the jet transverse energy distribution to match the data.
 
Certain simulation configurations are common to all samples.
The decays of bottom and charm hadrons were performed by \textsc{EvtGen}~\cite{Lange:2001uf}.
Minimum-bias events were simulated using the \PYTHIA 8.210 generator with the NNPDF2.3LO PDF set and the A3 parameter tune~\cite{ATL-PHYS-PUB-2016-017}.
A number of these events, varying in accord with the luminosity profile of the recorded data, were overlaid on the hard-scatter interactions to model pile-up contributions from both the same bunch crossing and neighbouring bunch crossings.
The response of the ATLAS detector to the generated events was then modelled using a full simulation of the detector~\cite{SOFT-2010-01} based on $\GEANT4$~\cite{Agostinelli:2002hh}.
\section{Object selection}
The event selection builds on standard ATLAS reconstruction algorithms for jets, photons, and leptons.
The leptons -- electrons and muons -- are identified only for the purpose of vetoing events with leptons to preserve orthogonality with other ATLAS measurements.
 
Jets are reconstructed from three-dimensional positive-energy topological clusters of calorimeter energy deposits calibrated to the electromagnetic scale~\cite{PERF-2014-07}.
These clusters are inputs to the \antikt jet reconstruction algorithm~\cite{Cacciari:2008gp,Fastjet} with a radius parameter of \(R=0.4\).
To suppress jets originating from pile-up vertices, a likelihood-based jet vertex tagger (JVT)~\cite{ATLAS-CONF-2014-018} is applied to jets with transverse momenta \(\pT < \SI{120}{\GeV}\) and \(|\eta|<2.5\).
A pile-up subtraction algorithm further reduces pile-up contributions to the jet energies~\cite{JETM-2018-05}.
These jet energies are calibrated with MC-derived correction factors, including \(\eta\)-dependent calibrations to ensure consistent jet energy measurements in the central and forward regions of the experiment~\cite{JETM-2018-05}.
The jet energies are further corrected using data-derived calibrations based on the \pT balance between jets and reference objects, such as \Zboson bosons, photons, or high-energy jets.
 
The MV2c10 multivariate \(b\)-tagging algorithm~\cite{FTAG-2018-01} tags \bjets\ (jets containing \(b\)-hadrons) within the tracker acceptance.
It uses log-likelihood ratios from two- and three-dimensional impact parameter distributions, secondary and tertiary vertex information, and the jet \pT and \(\eta\), as inputs to a BDT.
The requirement on the BDT output corresponds to a per-jet \(b\)-tagging efficiency of 77\%, as measured in simulated \ttbar events for \bjets with \(\pT > \SI{20}{\GeV}\) and \(|\eta|<2.5\), and \(c\)-jet and light-flavour-jet efficiencies of 25\% and 0.9\%, respectively.
Scale factors are applied to account for efficiency differences between simulated events and data~\cite{FTAG-2018-01,ATLAS-CONF-2018-001,ATLAS-CONF-2018-006}.
 
Photon and electron reconstruction begins with clusters of calorimeter energy deposits~\cite{EGAM-2018-01}.
Clusters without any matching track or conversion vertex are identified as unconverted photons, while clusters with a matching conversion vertex reconstructed from one or two tracks are identified as converted photons.
Clusters with a matching track, re-fitted to account for bremsstrahlung, are identified as electrons.
Calorimeter and track information for each photon or electron candidate is used to construct multivariate discriminants for identification, and `tight' selections are applied to both the photons and electrons.
To suppress hadronic background, further isolation requirements are optimized with simulated \(t\bar t\) events~\cite{EGAM-2018-01}.
For photon candidates, the calorimeter isolation variable \(\ET^\text{iso}\) is the sum of the transverse energies of topological clusters reconstructed in the electromagnetic and hadronic calorimeters in a cone of size \(\Delta R = 0.4\) around the photon candidate,
where the \(\Delta\eta\times\Delta\phi\) region of size \(0.125\times 0.175\) around the photon cluster's centroid is excluded.
The isolation requirement, which depends explicitly on the photon transverse energy \(\ET^\gamma\), is \(\ET^\text{iso} < \SI{2.45}{\GeV} + 0.022 \times \ET^\gamma\).
For electron candidates, both track- and calorimeter-based isolation is required.
The track-based isolation requirement is a function of the electron transverse momenum \(\pT^e\) and is based on the tracks within a cone of \(\pT^e\)-dependent size up to \(\Delta R = 0.2\) around the electron candidate~\cite{EGAM-2018-01}.
The calorimeter-based isolation requires that the sum of cluster transverse energies within the same \(\Delta R\) be less than \SI{3.5}{\GeV}.
 
Muons are reconstructed by combining inner detector tracks, where available, and muon spectrometer tracks up to \(|\eta|=2.7\).
They must satisfy the `loose' identification criteria~\cite{PERF-2015-10}.
Identified muons must be isolated from other tracks, with a total summed track \pT less than \SI{1.25}{\GeV} in a cone of \(\Delta R=0.2\) around the muon.
Only `loose' muons within the coverage of the inner detector, \(|\eta|<2.5\), are used in this selection.
 
Overlap between identified photons, leptons, and jets is removed with the following procedure.
First, photons within \(\Delta R = 0.4\) of any muon or electron are removed.
Next, jets within \(\Delta R = 0.2\) of any electron are removed, and electrons within \(\Delta R = 0.4\) of any remaining jets are removed.
Then, jets within \(\Delta R = 0.2\) of any muon are removed if fewer than three tracks are associated with the jet or if both \(\pt^\mu>0.5 \pt^\text{jet}\) and \(\pt^\mu>0.7 \sum \pt^\text{track}\), where \(\sum \pt^\text{track}\) is the sum of track transverse momenta associated with the jet.
Muons within \(\Delta R = \min(0.4, 0.04+\SI{10}{\GeV}/\pt^\mu)\) of any remaining jets are removed.
Finally, jets within \(\Delta R = 0.4\) of any photon are removed.
 
\section{Event selection}
The event selection criteria are based on the object reconstruction algorithms, with additional event-level requirements to select events compatible with VBF \(H \gamma\) production.
These criteria are similar to the event selection requirements in previous searches~\cite{HIGG-2016-30}.
 
The first-level trigger selection requires an isolated electromagnetic calorimeter object -- the photon -- with \(\ET > \SI{22}{\GeV}\).
The high-level trigger selects events in the specific VBF-enhanced phase space, defined by the following requirements.
An isolated reconstructed photon with \(\ET > \SI{25}{\GeV}\) is required in addition to at least four jets with \(\ET > \SI{35}{\GeV}\) and \(|\eta| < 4.9\).
The requirement that at least one pair of jets in the event has an invariant mass greater than \SI{700}{\GeV} targets the VBF phase space.
For most of Run 2, a \(b\)-tagging trigger algorithm was used to ensure that at least one jet was \(b\)-tagged at the 77\% efficiency working point~\cite{ATL-PHYS-PUB-2015-022}.
 
Additional selection requirements are placed on events that pass the trigger selection.
At least one photon with \(\ET > \SI{30}{\GeV}\) in the regions \(|\eta|<1.37\) or \(1.52 < |\eta| < 2.37\) must match the trigger photon.
Events must have at least four jets satisfying \(\pT > \SI{40}{\GeV}\) and \(|\eta|<4.5\), with at least two jets in \(|\eta|<2.5\) passing the \(b\)-tagging selection.
The two highest-\pT \(b\)-tagged jets are assumed to be from the Higgs boson decay; they are identified as \(b1\) and \(b2\) and at least one of them must match a \(b\)-tagged trigger-level jet when the \(b\)-tagging trigger algorithm is used.
Dedicated \bjet energy corrections are applied to \(b\)-tagged jets to improve their energy measurement (scale and resolution)~\cite{Aad:2020jym}.
They equalize the response to jets with semileptonic or hadronic decays of heavy-flavour hadrons and correct for resolution effects.
This correction improves the di-\bjet invariant mass \(m_{bb}\) resolution by 10\%.
From the remaining jets, the jet pair with the highest invariant mass is chosen as the VBF jets; they are identified as \(j1\) and \(j2\).
Requiring this invariant mass \(m_{jj}\) to be greater than \SI{800}{\GeV} ensures full efficiency for the trigger \(m_{jj}\) requirement.
Events containing identified and isolated electrons with \(\pT>\SI{25}{\GeV}\) and \(|\eta|<2.47\) or muons with \(\pT>\SI{25}{\GeV}\) are vetoed to avoid overlap with other Higgs boson event selections in ATLAS.
 
The jet \pT requirements in the trigger algorithms and offline event selection can introduce concavity in the \(m_{bb}\) distribution, making it more difficult to parameterize with an analytic function.
The concavity is removed by requiring the \pT of the di-\bjet system \(\pT(\bbbar)\) to be greater than \SI{60}{\GeV}, a value that was optimized in the large MC sample of non-resonant \(\bbbar \gamma jj\) background events.
 
The full list of trigger algorithm and offline event selection requirements, before the event-level BDT classification, is summarized in Table~\ref{tab:eventselection}.
 
\begin{table}[htbp]
\def\arraystretch{1.2}
\centering{
\begin{tabular}{ l | l | l }
\hline \hline
\multirow{4}{*}{Trigger} &     L1                    & \(\geq1\) photon with \(\ET > \SI{22}{\GeV}\) \\
\cline{2-3}
& \multirow{3}{*}{HLT}  & \(\geq1\) photon with \(\ET > \SI{25}{\GeV}\) \\
&         & \(\geq4\) jets (or \(\geq3\) jets and \(\geq1\) \bjet) with \(\ET>\SI{35}{\GeV}\) and \(|\eta|<4.9\)  \\
&                         & \(m_{jj}>\SI{700}{\GeV}\) \\
\hline
\multicolumn{2}{c|}{\multirow{5}{*}{Offline}} & \(\geq1\) photon with \(\ET>\SI{30}{\GeV}\) and \(|\eta|<1.37\) or \(1.52<|\eta|<2.37\) \\
\multicolumn{2}{c|}{}                         & \(\geq2\) \bjets with \(\pT>\SI{40}{\GeV}\) and \(|\eta|<2.5\) \\
\multicolumn{2}{c|}{}                         & \(\geq2\) jets with \(\pT>\SI{40}{\GeV}\) and \(|\eta|<4.5\) \\
\multicolumn{2}{c|}{}                         & \(m_{jj} > \SI{800}{\GeV}\) \\
\multicolumn{2}{c|}{}                        & \(\pT(\bbbar)>\SI{60}{\GeV}\) \\
\multicolumn{2}{c|}{}                        & No electrons (\(\pT>\SI{25}{\GeV}, |\eta|<2.47)\) or muons (\(\pT>\SI{25}{\GeV}, |\eta|<2.5)\) \\
\hline \hline
\end{tabular}
\caption{Trigger and offline event selection criteria for the \(H(\rightarrow b\bar{b})\gamma jj\) signature.
L1 and HLT refer to the first-level trigger and the high-level trigger, respectively.
\label{tab:eventselection}
}\par
}
\end{table}
\section{Multivariate analysis}
An event-level BDT classifies events as being signal-like or background-like, based on a set of kinematic variables selected to optimize the separation.
The input variables are chosen to ensure the BDT output discriminant shows low correlation with with \(m_{bb}\) to prevent distortions of the \(m_{bb}\) spectrum.
The following variables, ordered by decreasing classification power~\cite{TMVA}, are used for the BDT training:
\begin{enumerate}
\item \(\Delta \eta (jj)\), the pseudorapidity difference between the two VBF jets;
\item \(\pT^\text{balance}\), the transverse momentum balance for selected final-state objects, defined as
\begin{equation}\nonumber
\pt^\text{balance}=\frac{ |\vec{\pt}^{b1} +\vec{\pt}^{b2}+\vec{\pt}^{j1} +\vec{\pt}^{j2} +\vec{\pt}^{\gamma} |}{ \pt^{b1} +\pt^{b2}+\pt^{j1} +\pt^{j2} +\pt^{\gamma}};
\end{equation}
\item \(\Delta R (b1, \gamma)\), the angular distance between the leading \bjet and the photon;
\item \(m_{jj}\), the invariant mass of the two VBF jets;
\item \(\Delta R (b2, \gamma)\), the angular distance between the subleading \bjet and the photon;
\item \(\text{centrality}(\gamma, j1, j2)\), the rapidity of the photon relative to the VBF jet rapidities, defined as
\begin{equation}\nonumber
\text{centrality}(\gamma, j1, j2)=\left|\frac{y_{\gamma}-\frac{y_{j1}+y_{j2}}{2}}{y_{j1}-y_{j2}}\right|;
\end{equation}
\item \(\Delta \phi(\bbbar, jj)\), the azimuthal angle difference between the di-\bjet system and the VBF jet system;
\item \(\pT^{jj}\), the transverse momentum of the VBF jets system;
\item \(\cos \theta\), the cosine of the angle between the VBF jets plane and \bjets plane in the centre-of-mass frame of the \(\bbbar jj\) system;
\footnote{This angular variable is sensitive to the spin of the particle decaying into \bbbar. It helps discriminate Higgs boson production from gluon splitting~\cite{CMS-HIG-14-004}.} and
\item \(\Delta R(b1,j1)\), the angular distance between the leading \bjet and the leading VBF jet.
\end{enumerate}
 
\begin{figure}[h]
\begin{center}
\includegraphics[width=0.49\textwidth]{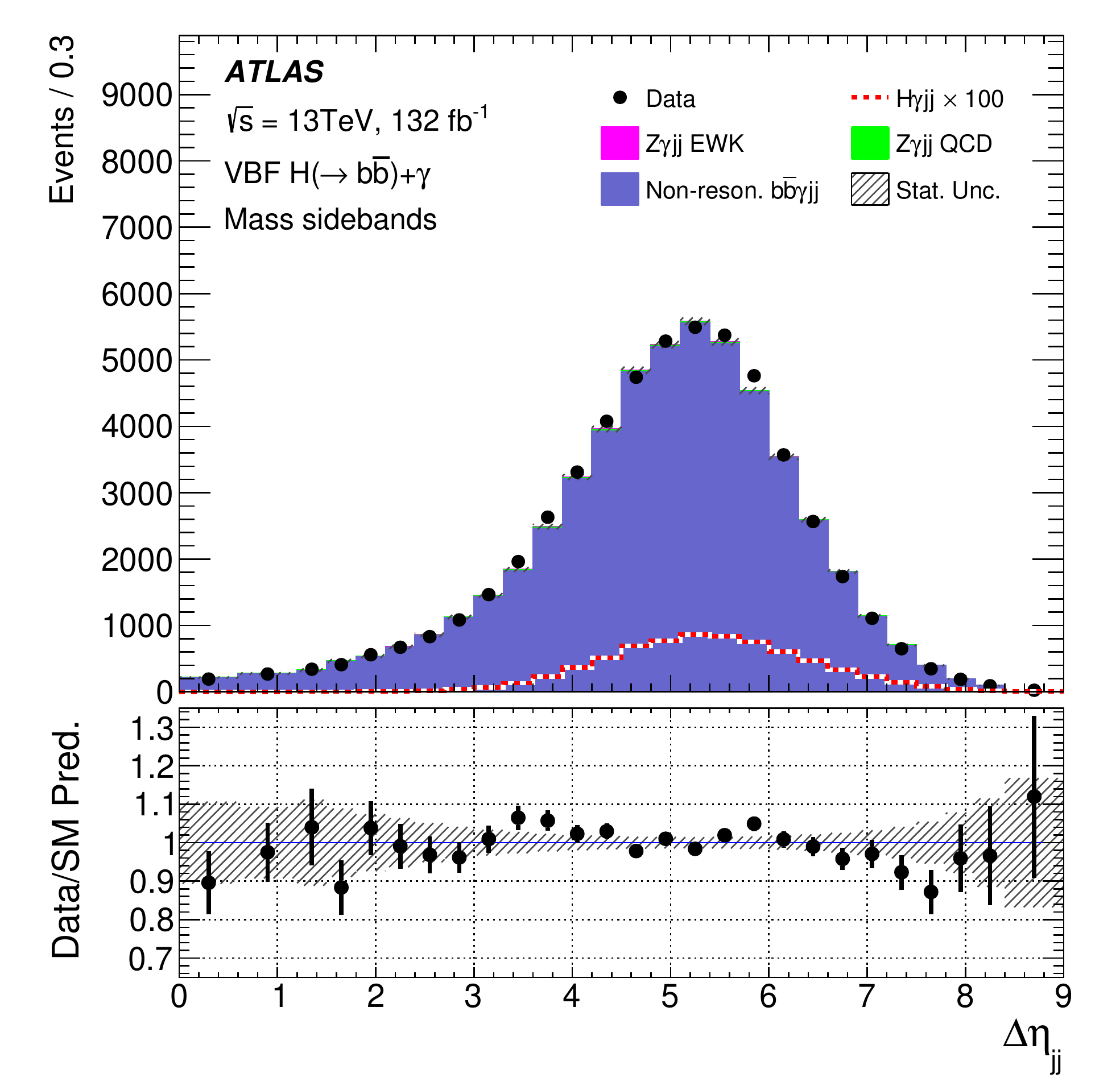}
\includegraphics[width=0.49\textwidth]{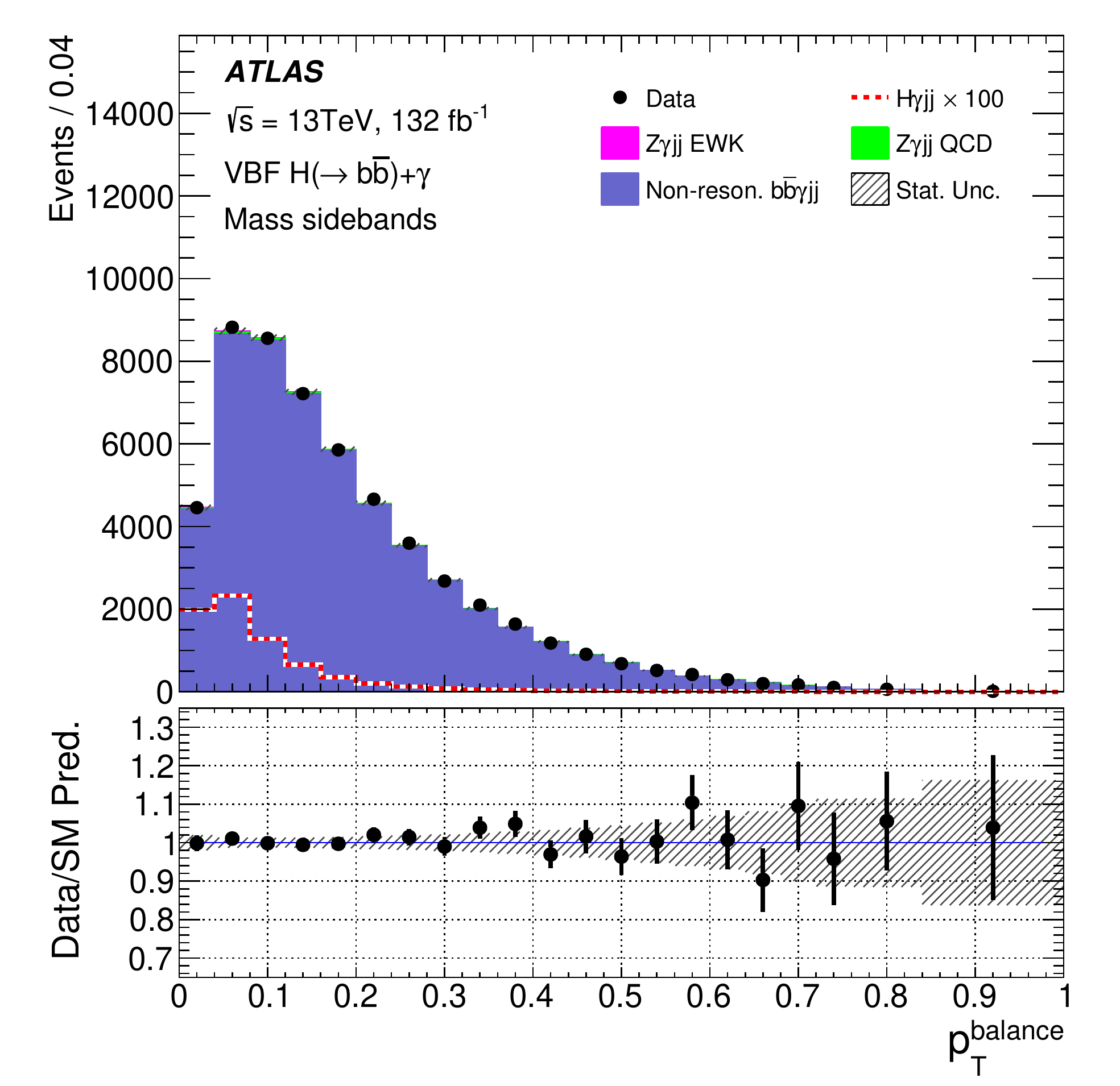}
\caption{Comparisons of data and simulated event distributions of the BDT input variables \(\Delta \eta_{jj}\) (left) and \(\pt^\text{balance}\) (right) in the two \(m_{bb}\) sidebands after kinematic reweighting of the non-resonant \(\bbbar\gamma jj\) background.
The data are shown as black points, and the background contributions are stacked in coloured histograms.
The Higgs boson signal contribution is scaled up and represented by the dashed red line.
The bottom panel in each plot shows the ratio of the data to the SM prediction, where the uncertainty band corresponds to the statistical uncertainty only.
\label{fig:bdtinputs}
}
\end{center}
\end{figure}

The \(H\gamma jj\) signal sample and the non-resonant \(\bbbar \gamma jj\) background sample are used for the BDT training in the TMVA package~\cite{TMVA}.
To improve agreement between the LO non-resonant \(\bbbar \gamma jj\) simulation and data,
analytic correction functions are fit in the mass sidebands (\(m_{bb} < \SI{100}{\GeV}\) and \(m_{bb}>\SI{140}{\GeV}\)). They are
based on data-to-MC ratios of the distributions for several relevant observables and are used to reweight the simulated events.
The overall normalization is also corrected to match the data in the mass sidebands.
This procedure is applied sequentially to the distributions of \(\Delta\eta_{jj}, \min[\Delta R(b1,\gamma), \Delta R(b2, \gamma)]\), and \(\pt^\text{balance}\), resulting in corrections that are typically less than 10\%.
The distributions of the other uncorrelated input variables are not significantly affected by the reweighting.
Comparisons of the data and MC distributions for the two most powerful classification variables, after the reweighting procedure, are shown in Figure~\ref{fig:bdtinputs}.
 
\begin{figure}[tbph]
\begin{center}
\includegraphics[width=0.8\textwidth]{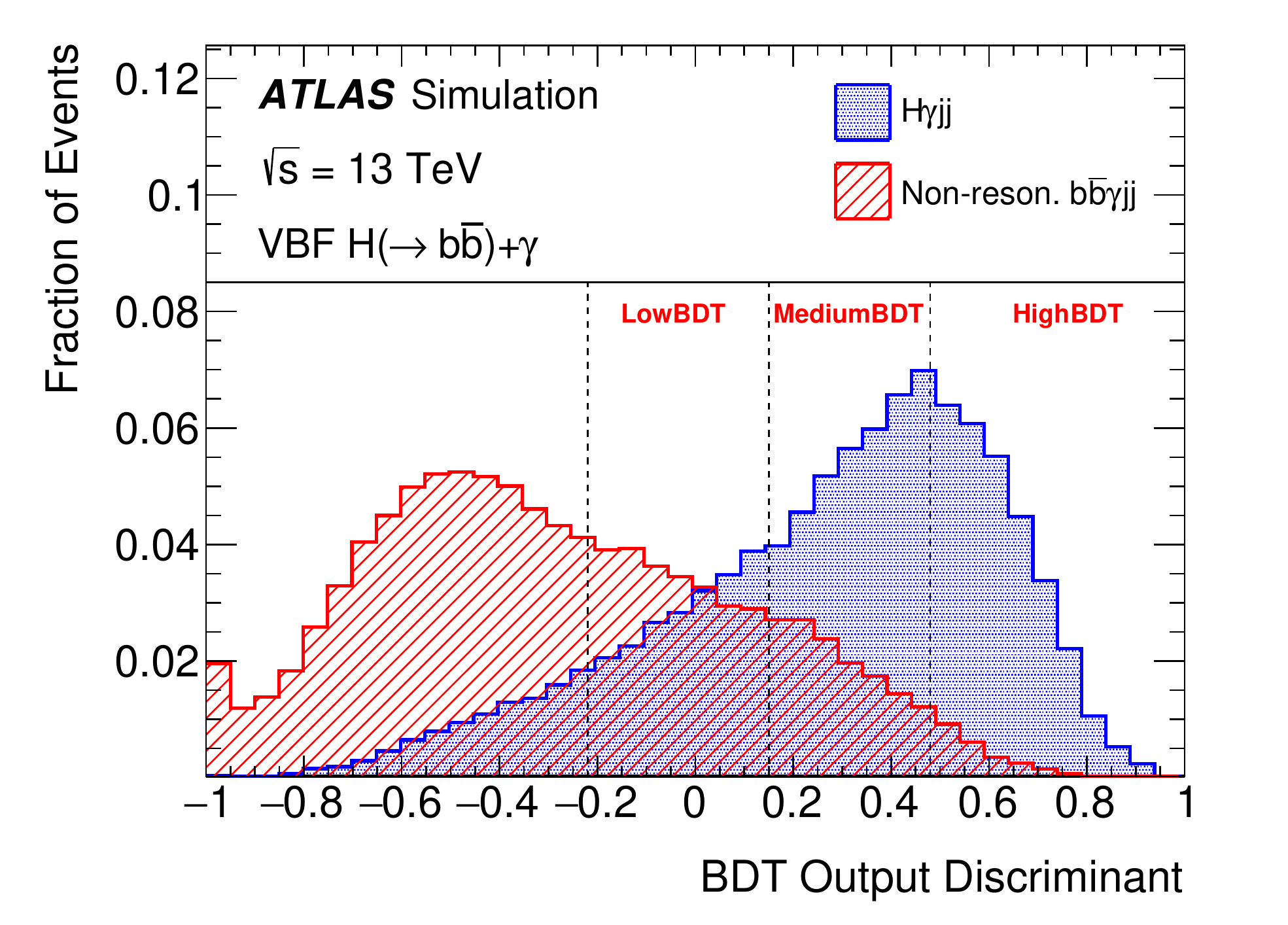}
\end{center}
\caption{BDT output discriminant distributions for the \(H\gamma jj\) signal and non-resonant \(\bbbar \gamma jj\) background samples.
The distributions from the signal and background samples are shown as blue and red hatched histograms, respectively.
The vertical dashed lines represent the boundaries of the event categories, defined by the BDT output discriminant.
\label{fig:bdtoutput}
}
\end{figure}

The BDT output discriminant is used to define three signal categories: HighBDT, MediumBDT, and LowBDT.
The boundaries of the three categories are defined sequentially from HighBDT to LowBDT by maximizing the combined VBF Higgs boson signal significance across categories.
The BDT output distributions for the signal and background samples within the three categories are shown in Figure~\ref{fig:bdtoutput}.
 
\section{Signal and background modelling}
\label{sec:modelling}
The main sources of background contributing to the \(\bbbar \gamma jj\) final-state signature are divided into processes with the decay of a massive gauge boson into \(b\)-tagged jet pairs and processes with non-resonant \(b\)-tagged jet pairs.
The resonant background is dominated by \(Z(\rightarrow \bbbar)\gamma jj \), with a negligible contribution from \(W\gamma jj\).
The non-resonant background is dominated by multijet \(\bbbar \gamma jj\) production, with small
contributions from \(\ttbar\gamma\) and single-top events.
The expected contributions from the dominant signal and background processes are collected in Table~\ref{tab:yields}.
 
\begin{table}[tbhp]
\renewcommand\arraystretch{1.2}
\begin{center}
\begin{tabular}{l r @{\hspace{.1\tabcolsep}} l @{\hspace{.4\tabcolsep}}  l r @{\hspace{.1\tabcolsep}} l @{\hspace{.4\tabcolsep}} l r @{\hspace{.1\tabcolsep}} l @{\hspace{.4\tabcolsep}} l}
\hline
\hline
Category &  \multicolumn{3}{c}{LowBDT} &  \multicolumn{3}{c}{MediumBDT} &   \multicolumn{3}{c}{HighBDT}\\
\hline
BDT Output      & \multicolumn{3}{c}{\((-0.22, 0.15)\)}      & \multicolumn{3}{c}{\((0.15, 0.48)\)}        & \multicolumn{3}{c}{\((0.48, 1)\)}           \\
\hline
VBF+\(VH\)    &   \(11\)&\(.31\) & \(\pm\) \hspace{3pt} 0.12    &   21 & .07&\(\pm\) \hspace{3pt} 0.17 &   18 & .33 & \(\pm\) \hspace{3pt} 0.15  \\
ggF     &   \(1\)& \(.71\) & \(\pm\) \hspace{3pt} 0.41    &   1 & .08  &\(\pm\) \hspace{3pt} 0.33    &    0 & .19 & \(\pm\) \hspace{3pt} 0.13  \\
\(\ttbar H\)     &   0 & .75& \(\pm\) \hspace{3pt} 0.14    &   0 & .19  & \(\pm\) \hspace{3pt}  0.03    &   0 & .05& \(\pm\) \hspace{3pt} 0.01  \\
\hline
\(\Zboson\gamma jj\) EWK     &   8 & .50& \(\pm\) \hspace{3pt} 0.14    &   8 & .91& \(\pm\) \hspace{3pt} 0.14    &   5& .37& \(\pm\) \hspace{3pt} 0.11  \\
\(\Zboson\gamma jj\) QCD      &   15 &.9 & \(\pm\) \hspace{3pt} 1.7    &   7&.2 \hspace{3pt} & \(\pm\) \hspace{3pt} 1.1    &   1& .65& \(\pm\) \hspace{3pt} 0.52  \\
Non-resonant \(\bbbar \gamma jj\)     & 2834 & & \(\pm\) 42\hspace{3pt}    &   1507& & \(\pm\) 33\hspace{3pt}    &   322& &\(\pm\) 17  \\
\hline
Expected Yield & 2872& &\(\pm\) 42 & 1545& &\(\pm\) 33 & 348& &\(\pm\) 17 \\
Data    &   2964 & & &   1522 & & &   318 & &   \\
\hline
\hline
\end{tabular}
\end{center}
\caption{Data and expected yields for the Higgs boson signal, resonant \(Z\gamma jj\) and non-resonant \(\bbbar \gamma jj\) background processes, in the three categories defined by the BDT output discriminant.
The event yields are calculated from simulated samples in the mass range \( \SI{100}{\GeV} < m_{bb} < \SI{140}{\GeV}\).
The yields are shown with statistical uncertainties only because experimental and theoretical systematic uncertainties, which depend on fits to data, are an order of magnitude smaller.
\label{tab:yields}
}
\end{table}

To extract the Higgs boson production rate from fits to data, the signal and dominant backgrounds -- the resonant \(Z\gamma jj\) and non-resonant \(\bbbar \gamma jj\) contributions -- are parameterized with analytic functions derived from simulated events or data sideband regions.
The small contributions from other backgrounds are included in the non-resonant background functions.
 
\begin{figure}
\includegraphics[width=0.49\textwidth]{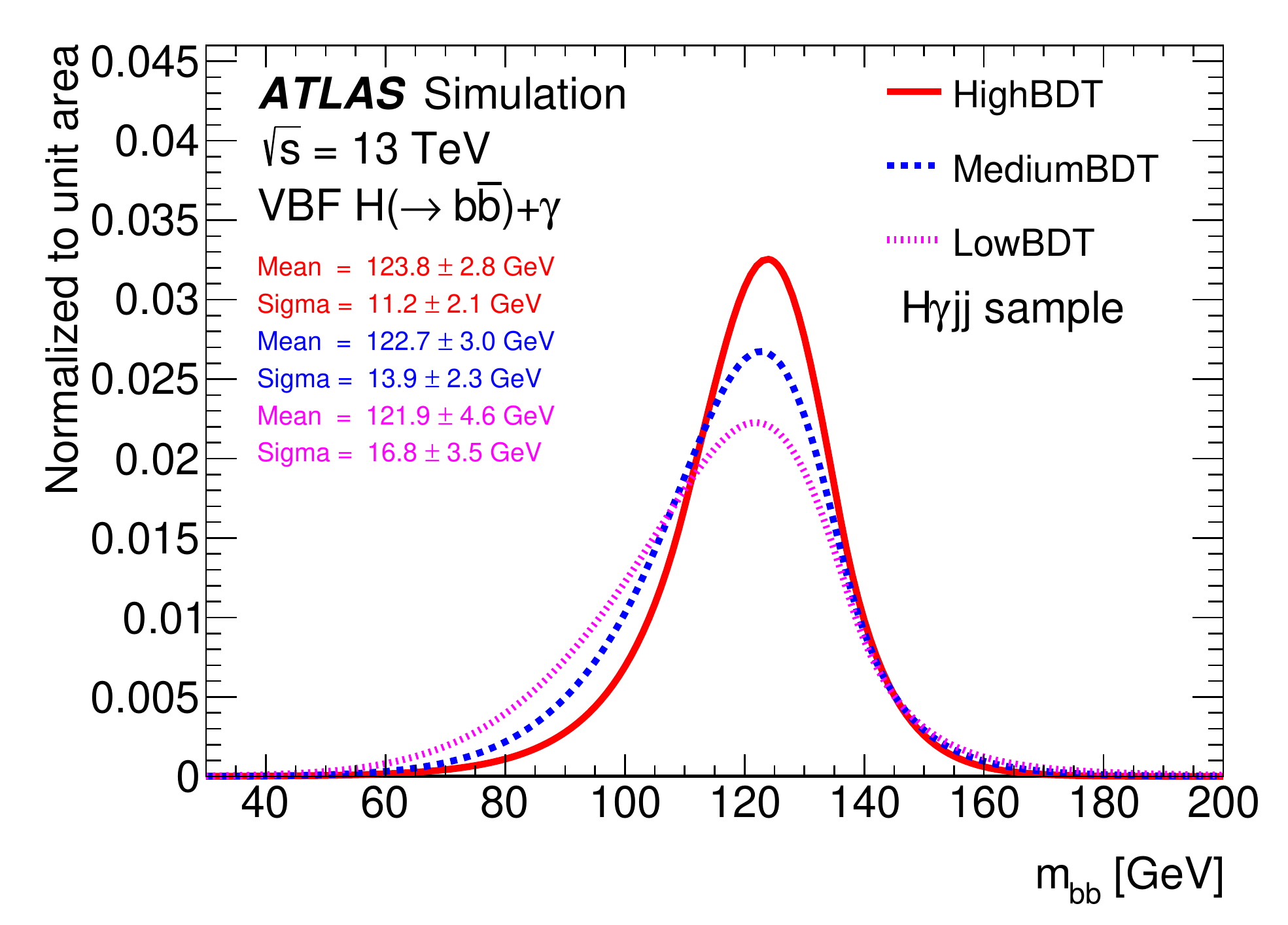}
\includegraphics[width=0.49\textwidth]{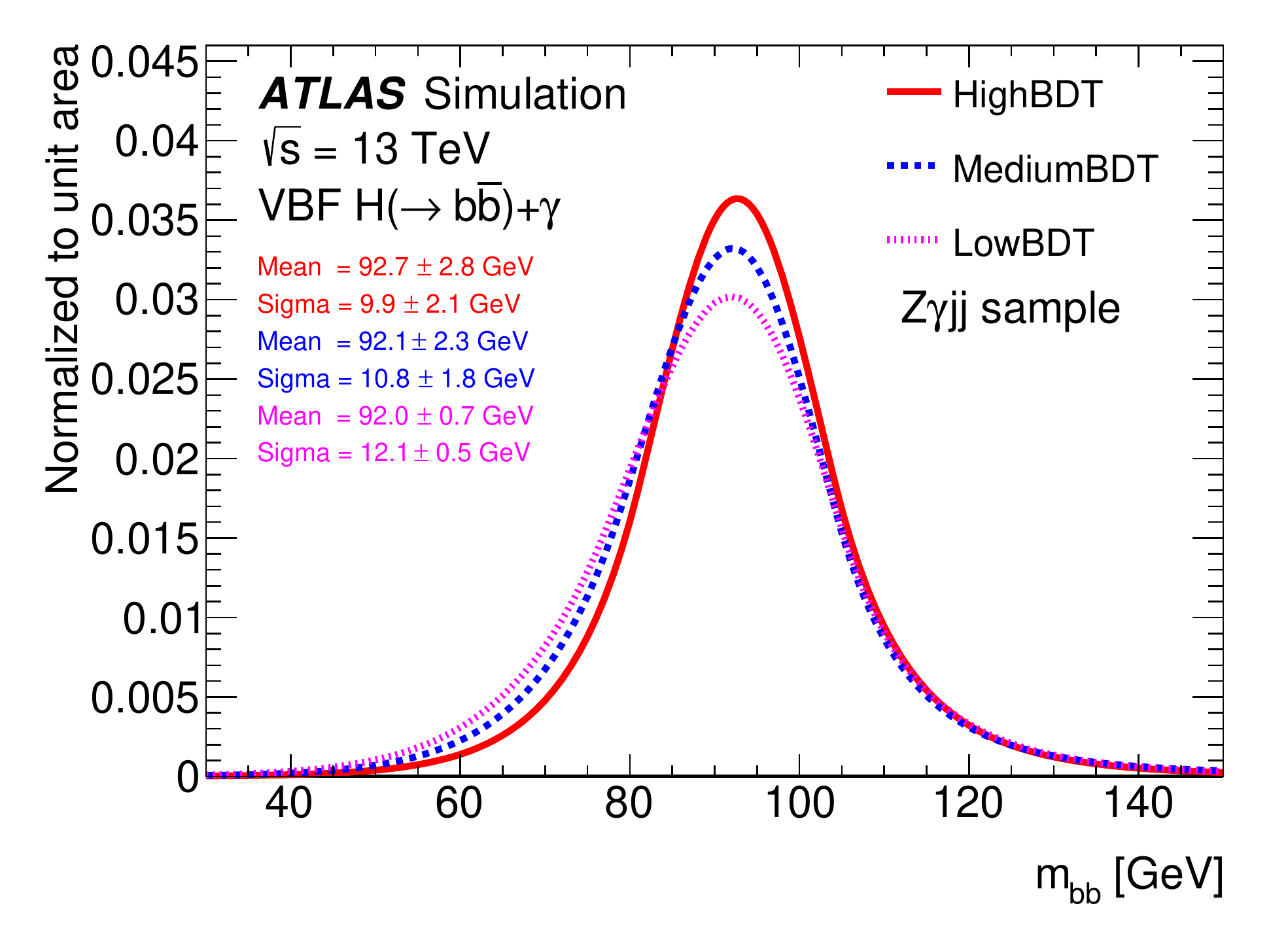}
\caption{Comparison of the Bukin function parameterizations, normalized to unit area, of the \(m_{bb}\) distributions for \(H\gamma jj\) events (left) and \(Z \gamma jj\) events (right) in the three BDT categories, along with the mean and sigma values for the three fitted functions.
\label{fig:param_compare}
}
\end{figure}

The Higgs boson signal \(m_{bb}\) distribution is modelled with a Bukin function~\cite{bukin2007fitting}, parameterized using the simulated samples independently in each of the three event categories defined by the BDT output, as shown in Figure~\ref{fig:param_compare}.
Among the possible functions, the Bukin function showed the best fit quality, as gauged by fit residuals across the mass range.
With this function, the residuals follow a normal distribution about zero.
The resonant \(Z\gamma jj\) background is also described with the same Bukin function using independent parameters in each of the three BDT categories.
 
The model for the non-resonant background is tested and largely determined using the data sidebands outside the mass window \(\SI{100}{\GeV} < m_{bb} < \SI{140}{\GeV}\).
Several different classes of fitting functions were tested, including polynomial, Bernstein polynomial and power-law functions.
The potential bias arising from the fitting procedure, termed `spurious signal', is estimated by performing signal-plus-background fits over the full mass range in the large simulated sample of non-resonant \(\bbbar \gamma jj\) background events.
The polynomial function, with two terms (first-order) for the HighBDT region or three terms (second-order) for the MediumBDT and LowBDT regions, has the smallest bias among the tested functions and is therefore chosen for the background template in the fit.
The standard \(F\)-test is used as a confirmation check, assessing the significance of decreases in the reduced \(\chi^2\) value as the order of the background-only polynomial fit to the data sidebands is increased.
The spurious-signal uncertainty, evaluated separately in each BDT region as the maximum of the fitted spurious signal and the MC statistical uncertainty of the non-resonant \(\bbbar \gamma jj\) background, is included as a systematic uncertainty in the signal yield.
The spurious signal is treated in the same way when \(Z\gamma jj\) is treated as signal.
 
\section{Systematic uncertainties}
Systematic uncertainties in the Higgs boson production rate arise from choices of modelling and fitting methods -- such as the spurious-signal uncertainties -- and from uncertainties in the event reconstruction or cross-section calculations.
They are summarized in Table~\ref{tab:breakdown} together with the statistical uncertainties calculated using the methods described in Section~\ref{sec:results}.
The systematic uncertainties for the background and signal contributions are divided into experimental and theoretical uncertainties.
They affect only the signal and background predictions that are based on simulation; they do not affect the non-resonant background estimate, which is derived from the data only.
The uncertainties are propagated through the event selection and the BDT classification to the template distributions.
They modify the shape and nominal normalization of the Higgs boson signal process. However, they modify only the shape of the resonant \(\Zboson\gamma\) background, as its normalization is derived from the \(m_{bb}\) fits in each category separately.
In general, the impact of the signal shape uncertainties on the final Higgs boson measurement is smaller than the impact of the experimental systematic uncertainties.
 
Other uncertainties are related to factors in the full \(m_{bb}\) fit, such as the uncertainty in the \(\Zboson\gamma\) normalizations (`\Zboson boson normalizations' in Table~\ref{tab:breakdown}).
The uncertainties in the \(\bbbar \gamma jj\) background relate to the fitted shapes (`Bkg.\ fit shapes') and normalizations (`Bkg.\ fit normalizations').
 
\subsection{Experimental uncertainties}
The uncertainty in the combined 2015--2018 integrated luminosity is 1.7\%~\cite{ATLAS-CONF-2019-021}, obtained using the LUCID-2 detector~\cite{LUCID2} for the primary luminosity measurements.
This uncertainty is applied to all contributions estimated from simulated samples alone.
 
The dominant jet-related uncertainties for the \(m_{bb}\) reconstruction are the jet energy scale (JES) and jet energy resolution (JER) uncertainties.
These are determined by \pT-balancing methods in data~\cite{JETM-2018-05}, and the effect on the mass spectrum is determined by shifting or smearing the jet energies before calculating \(m_{bb}\).
A total of 8 JER and 30 JES uncertainty parameters are considered in calculating the effects.
The systematic uncertainty in the JVT is estimated by varying the tagger efficiency within its uncertainties.
All of these jet uncertainties are summarized in the `Jet' line of Table~\ref{tab:breakdown}.
 
The uncertainties related to \(b\)-tagging jets, covering the trigger-level \(b\)-tagging efficiencies and the offline \(b\)-tagging efficiency data-to-simulation scale factors, are implemented as variations of the event-weight scale factors.
They are determined from data using \ttbar events, \Wboson boson decays into hadrons, and multijet data~\cite{FTAG-2018-01,ATLAS-CONF-2018-001,ATLAS-CONF-2018-006}.
The event weight is calculated from the product of the \(b\)-tagging scale factors for the two \bjets in the event.
 
Additional uncertainties are related to the photon energy measurement and reconstruction efficiency~\cite{EGAM-2018-01}.
The measurement is only weakly sensitive to the photon energy; therefore a simplified two-parameter model is used to capture the effect of variations in the energy scale and resolution.
The efficiency variations considered are due to electromagnetic shower shape variations and isolation calculations.
Systematic uncertainties from electron and muon selections are negligible and are therefore ignored.
 
\subsection{Theoretical uncertainties}
 
The theoretical uncertainties on the SM calculations of the total Higgs boson production rates are relevant in the determination of the relative rate of \(H\gamma jj\) events compared to the SM expectation.
The \(H\rightarrow\bbbar\) branching ratio uncertainty follows the recommendation of the LHC Higgs Cross Section Working Group, including the uncertainty in the \bquark mass~\cite{deFlorian:2016spz}.
Cross-section uncertainties due to the choice of renormalization and factorization scales are estimated by varying the choice of both scales up and down by factors of two independently.
The largest variation in each BDT classification category is used to define the uncertainty.
Similar uncertainties are calculated for the set of PDF variations defined by the eigenvectors of the PDF4LHC15\_nlo\_mc\_pdfas set.
The impact on signal normalization is evaluated at the reconstruction level after the event selection, and an overall uncertainty is derived following the PDF4LHC recommendation~\cite{Butterworth:2015oua}.
Uncertainties in the \HERWIG7 parton shower are estimated by varying the `HardScaleFactor' parameter by factors of two and comparing the resulting acceptances~\cite{Bellm:2015jjp}.
\section{Fit results for Higgs boson production}
\label{sec:results}
The Higgs boson signal strength \(\mu_H \) is defined relative to the total SM prediction for \(H\gamma jj\) production, and the VBF signal strength \(\mu_\text{VBF}\) is defined relative to the VBF contribution only.
They are extracted from an unbinned extended maximum-likelihood fit to the di-\bjet invariant mass distribution in all three BDT classification categories.
The likelihood is defined as a product of global Poisson distributions with event-by-event probabilities determined from the signal-plus-background model and the constraints for systematic uncertainties, implemented as nuisance parameters.
The nuisance parameters control the effects of the systematic uncertainties and are parameterized by Gaussian or log-normal priors.
Each prior's definition constrains its nuisance parameter to the nominal value within its associated uncertainty.
 
The Higgs boson signal strength is the single parameter of interest, common to all three categories, while the \(\Zboson \gamma jj\) contribution strength is fit as three separate parameters, uncorrelated across the categories.
Using three \(\Zboson \gamma jj\) parameters allows for different relative contributions of QCD and EWK processes in the three categories.
The uncorrelated parameters describing the non-resonant \(\bbbar \gamma jj\) background are allowed to float during the fit to obtain the best independent description of the background in each category.
The experimental and theoretical uncertainties are correlated across categories during the fit, reflecting their common derivation and calibration.
Signal-injection tests, performed by adding simulated signal to the expected background, confirmed the linearity of the fit with no bias in \(\mu_H\).
 
\begin{figure}
\centering
\includegraphics[width=0.49\textwidth]{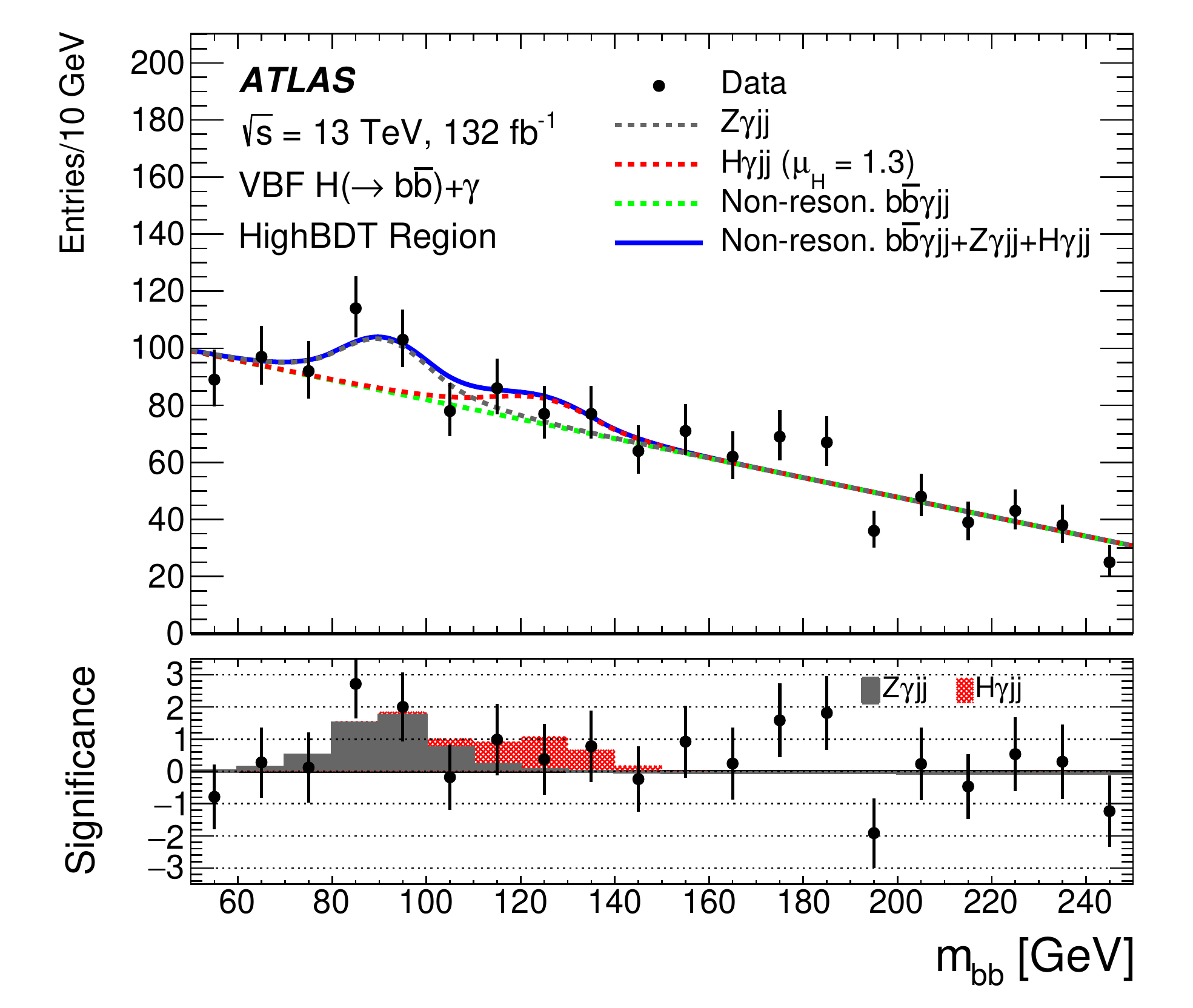}
\includegraphics[width=0.49\textwidth]{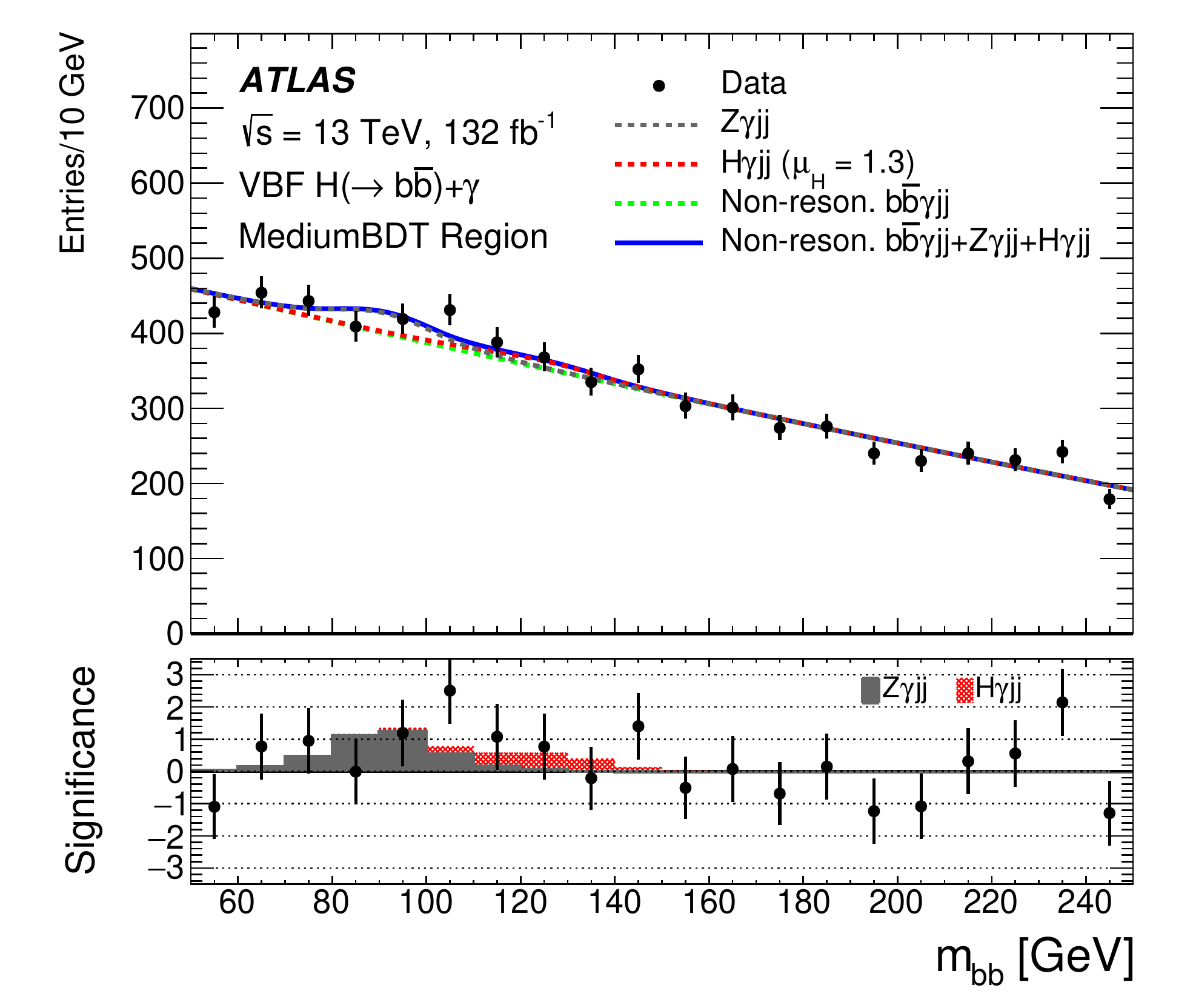}
\includegraphics[width=0.49\textwidth]{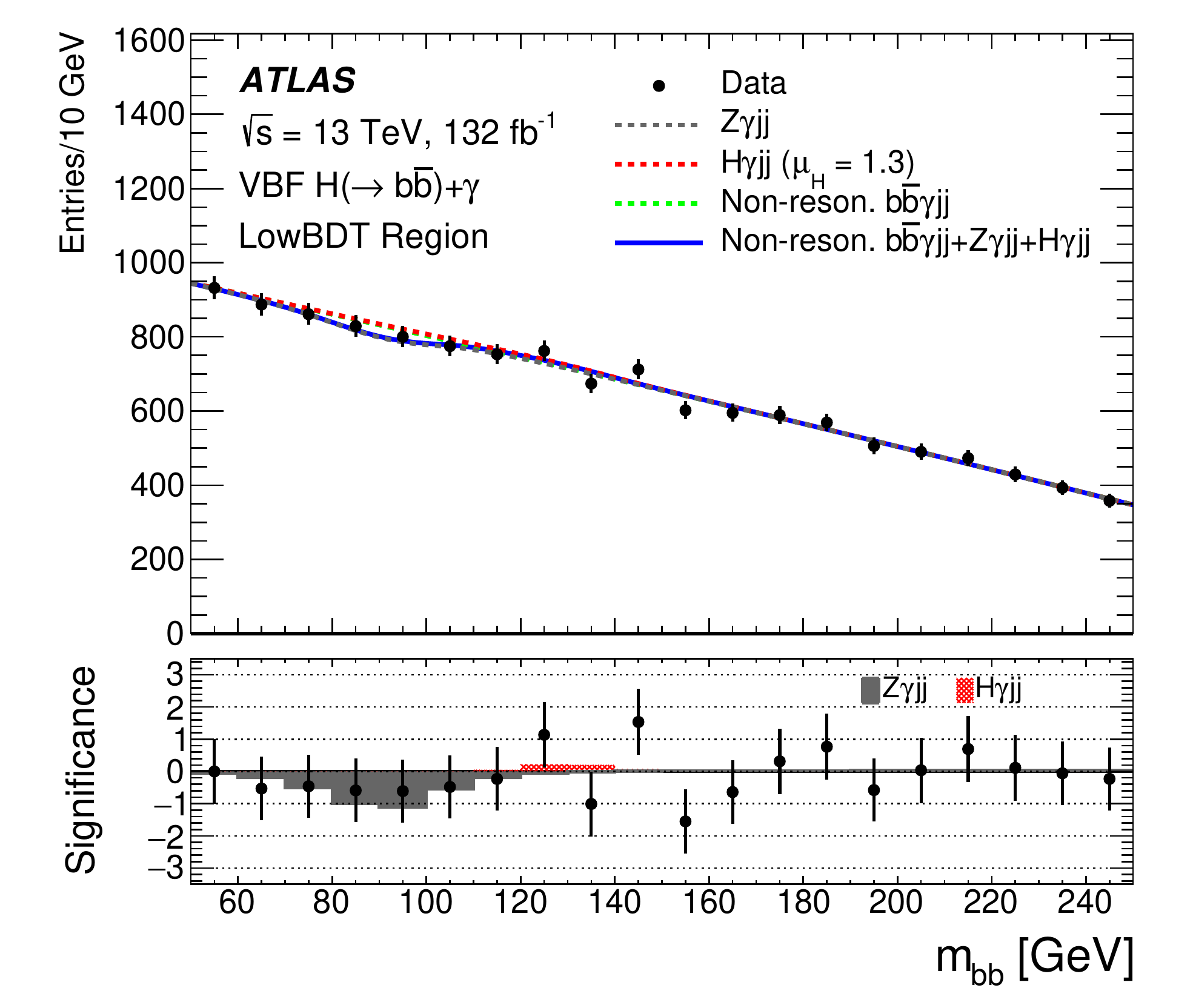}
\caption{The \(m_{bb}\) distributions in the three event categories, overlaid with contributions from the \(H\gamma jj\) signal as well as the resonant \(Z\gamma jj\) and non-resonant \(\bbbar \gamma jj\) background fits. The combined \(\chi^2\) per degree of freedom is \(45.2/45\). The bottom panel in each plot presents the significance of the Higgs boson signal relative to the non-resonant \(\bbbar \gamma jj\) background in each bin, calculated using a Poisson model~\cite{Choudalakis_2012}.
\label{fig:SR_distribution}
}
\end{figure}

The results of the fits to the \(m_{bb}\) distributions in data are shown in Figure~\ref{fig:SR_distribution}, with contributions from the Higgs boson, \Zboson boson, and background components superimposed.
The inclusive signal strength \(\mu_H\) and the VBF signal strength \(\mu_\text{VBF}\) are both \(1.3\pm 1.0\).
The similarity of the results is due to the nearly negligible contribution from other Higgs boson production modes in the VBF-enhanced phase space, defined by the high \(m_{jj}\) requirement.
If the inclusive signal strength is fit as three separate parameters of interest, the results are \(\mu_H = 0.7\pm 1.1, 3.8^{+2.5}_{-2.4}\), and \(3.8^{+7.0}_{-8.3}\) in the HighBDT, MediumBDT, and LowBDT categories, respectively.
 
The \Zboson normalization factors obtained from the fit are \(\mu_\Zboson = 1.9 \pm 1.2, 1.5 \pm 1.1\) and \(-1.3^{+1.2}_{-1.6}\) in the HighBDT, MediumBDT, and LowBDT categories, respectively.
The \Zboson normalization and the \(\mu_H\) signal strength have a correlation of 15\% in the HighBDT category but are uncorrelated in the other categories.
If the \Zboson normalization factors are constrained to be non-negative, the change to the combined Higgs boson signal strength \(\mu_H\) is within rounding errors.
 
The observed significance of the Higgs boson signal above the background is \(1.3\sigma\), compared to an expected significance of \(1.0\sigma\).
This measurement is consistent with previous results~\cite{HIGG-2016-30}.
The dominant uncertainty in the result is the statistical uncertainty of the limited data sample, followed by the background fit uncertainties and the spurious-signal uncertainty.
The relative contributions from those and other uncertainties are shown in Table~\ref{tab:breakdown}.
Impacts from individual contributions are estimated as the difference in quadrature between the Higgs signal strength uncertainties with each nuisance parameter floating and fixed.
The asymmetry observed in some of the systematic uncertainties is due to the limited measurement sensitivity: during the profile likelihood calculation, \(\mu_H\) can assume values close to 0.
 
\begin{figure}
\begin{center}
\includegraphics[width=0.55\textwidth]{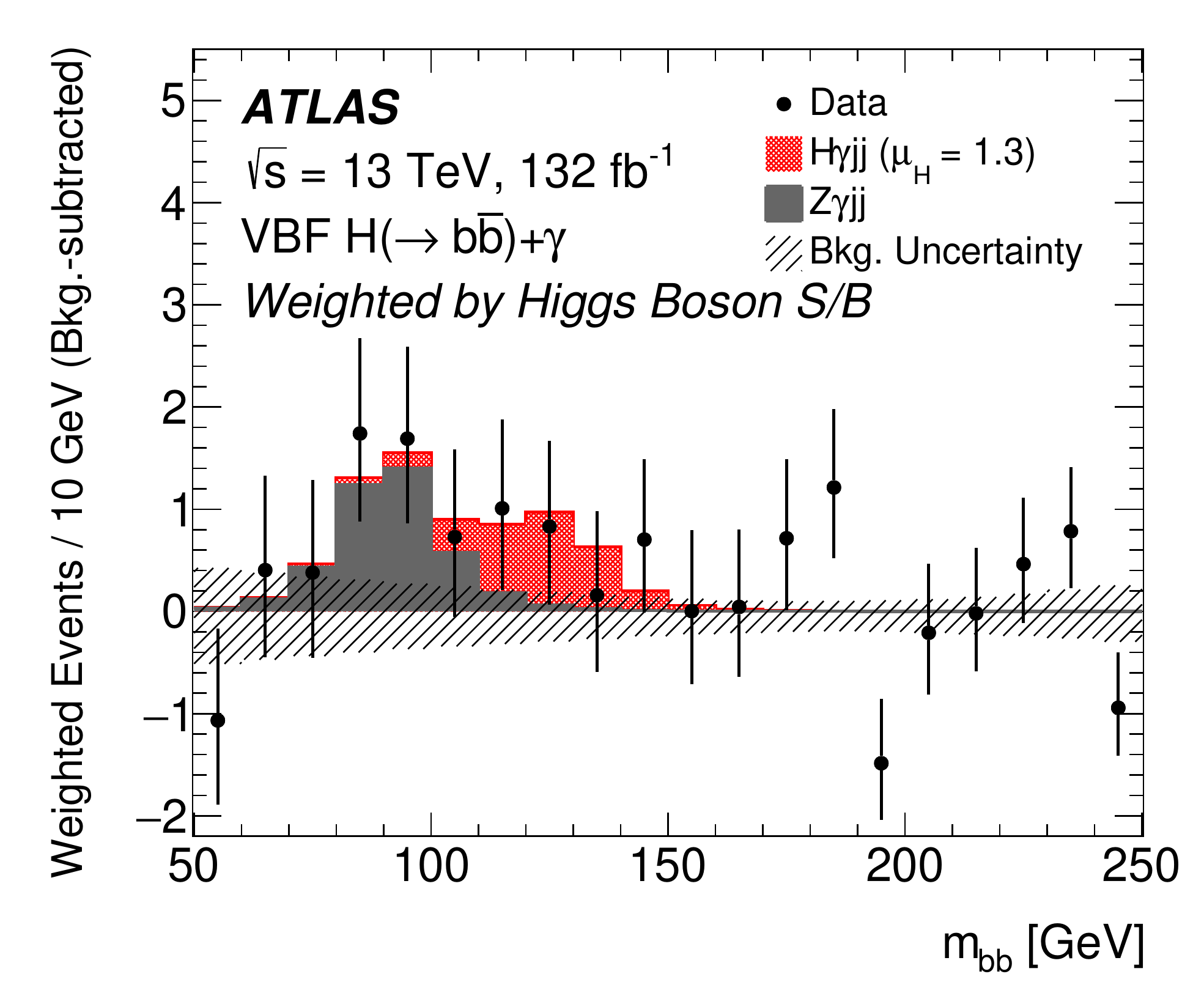}
\end{center}
\caption{The combined \(m_{bb}\) distribution, after subtracting the fitted non-resonant \(\bbbar \gamma jj\) background, summed over the three BDT categories weighted by \(S/B\) in each category, where \(S\) and \(B\) are the fitted Higgs boson signal yield and \(Z\gamma jj\) + \(\bbbar \gamma jj\) background yield, respectively, calculated in the mass window containing 68\% of the Higgs boson signal.
The expected contributions of \(H\gamma jj\) and \(Z\gamma jj\) are also shown, scaled by the fitted Higgs boson signal strength and the \(Z\gamma jj\) normalization factors in the three BDT categories.
The statistical uncertainty of the fitted non-resonant \(\bbbar\gamma jj\) background is represented by the hatched band.
The statistical uncertainty on data is represented by the error bars on the data points.
\label{fig:combined_mass}
}
\end{figure}

The combined mass distribution for the three BDT categories, after the non-resonant \(\bbbar \gamma jj\) background contribution has been subtracted, is shown in Figure~\ref{fig:combined_mass}.
The events in each BDT category have been weighted by the signal-to-background ratio \(S/B\), as calculated from the fitted signal and background contributions in the 68\% confidence-interval mass window around the Higgs boson signal peak.
 
\begin{table}
\begin{center}
\begin{tabular}{lcc}
\hline \hline
Source of absolute uncertainty &  \(\sigma(\mu_H)\) down & \(\sigma(\mu_H)\) up\\
\hline
Statistical\\
\quad Data statistical & \(-0.78\)  & \(+0.80\)\\
\quad Bkg.\ fit shapes &  \(-0.19\)  & \(+0.22\)\\
\quad Bkg.\ fit normalizations &  \(-0.51\)  & \(+0.52\)\\
\quad \Zboson boson normalizations & \(-0.15\)  & \(+0.14\)\\
\hline
Systematic\\
\quad Spurious signal & \(-0.24\)  & \(+0.21\)\\
\quad Theoretical &\(-0.01\)  & \(+0.08\)\\
\quad Photon &          \(-0.01\)  & \(+0.03\)\\
\quad Jet &             \(-0.06\)  & \(+0.20\)\\
\quad \(b\)-tagging &       \(-0.02\)  & \(+0.11\)\\
\quad Auxiliary &  \(-0.01\)  & \(+0.04\)\\
 
\hline \hline
Total & \(-0.99\) & \(+1.04\) \\
\quad Total statistical & \(-0.96\) & \(+0.99\) \\
\quad Total systematic & \(-0.25\) & \(+0.32\)\\
\hline \hline
\end{tabular}
\caption{The effect of the uncertainties on the signal strength.
The dominant contributions are from the statistical uncertainty of the dataset, background fit uncertainties, and the spurious-signal uncertainty.
The statistical uncertainty is calculated by fixing the background fit parameters to their nominal values.
Individual uncertainties are then combined to give the total within each group.
The small uncertainties from pile-up effects, luminosity measurements, and trigger scale factors are grouped under `Auxiliary' uncertainty.
\label{tab:breakdown}
}
\end{center}
\end{table}
 
\FloatBarrier
 
\section{Conclusion}
A search has been conducted for the SM Higgs boson produced in association with a high-energy photon in the \(H (\rightarrow \bbbar) \gamma j j\) signature, with a focus on the phase space typical of vector-boson fusion.
The search used the full LHC Run 2 dataset of proton--proton collisions at \(\sqrt{s}=\SI{13}{\TeV}\), corresponding to an integrated luminosity of \SI{132}{\ifb} collected with the ATLAS detector.
A BDT is used to separate events into categories, and the \(m_{bb}\) distribution is fit to extract the Higgs boson signal production rate.
The measured Higgs boson signal strength relative to the SM prediction is \(\mu_H = 1.3 \pm 1.0\).
This corresponds to an observed signal significance of 1.3 standard deviations, compared to an expected significance of 1.0 standard deviations.
The improvement over the previous measurement of \(\mu_H = 2.3 \pm 1.8\) comes from the larger dataset, the updated BDT, and more precise  Monte Carlo modelling.
 
\section*{Acknowledgements}
 

We thank CERN for the very successful operation of the LHC, as well as the
support staff from our institutions without whom ATLAS could not be
operated efficiently.
 
We acknowledge the support of ANPCyT, Argentina; YerPhI, Armenia; ARC, Australia; BMWFW and FWF, Austria; ANAS, Azerbaijan; SSTC, Belarus; CNPq and FAPESP, Brazil; NSERC, NRC and CFI, Canada; CERN; ANID, Chile; CAS, MOST and NSFC, China; COLCIENCIAS, Colombia; MSMT CR, MPO CR and VSC CR, Czech Republic; DNRF and DNSRC, Denmark; IN2P3-CNRS and CEA-DRF/IRFU, France; SRNSFG, Georgia; BMBF, HGF and MPG, Germany; GSRT, Greece; RGC and Hong Kong SAR, China; ISF and Benoziyo Center, Israel; INFN, Italy; MEXT and JSPS, Japan; CNRST, Morocco; NWO, Netherlands; RCN, Norway; MNiSW and NCN, Poland; FCT, Portugal; MNE/IFA, Romania; JINR; MES of Russia and NRC KI, Russian Federation; MESTD, Serbia; MSSR, Slovakia; ARRS and MIZ\v{S}, Slovenia; DST/NRF, South Africa; MICINN, Spain; SRC and Wallenberg Foundation, Sweden; SERI, SNSF and Cantons of Bern and Geneva, Switzerland; MOST, Taiwan; TAEK, Turkey; STFC, United Kingdom; DOE and NSF, United States of America. In addition, individual groups and members have received support from BCKDF, CANARIE, Compute Canada, CRC and IVADO, Canada; Beijing Municipal Science \& Technology Commission, China; COST, ERC, ERDF, Horizon 2020 and Marie Sk{\l}odowska-Curie Actions, European Union; Investissements d'Avenir Labex, Investissements d'Avenir Idex and ANR, France; DFG and AvH Foundation, Germany; Herakleitos, Thales and Aristeia programmes co-financed by EU-ESF and the Greek NSRF, Greece; BSF-NSF and GIF, Israel; La Caixa Banking Foundation, CERCA Programme Generalitat de Catalunya and PROMETEO and GenT Programmes Generalitat Valenciana, Spain; G\"{o}ran Gustafssons Stiftelse, Sweden; The Royal Society and Leverhulme Trust, United Kingdom.
 
The crucial computing support from all WLCG partners is acknowledged gratefully, in particular from CERN, the ATLAS Tier-1 facilities at TRIUMF (Canada), NDGF (Denmark, Norway, Sweden), CC-IN2P3 (France), KIT/GridKA (Germany), INFN-CNAF (Italy), NL-T1 (Netherlands), PIC (Spain), ASGC (Taiwan), RAL (UK) and BNL (USA), the Tier-2 facilities worldwide and large non-WLCG resource providers. Major contributors of computing resources are listed in Ref.~\cite{ATL-SOFT-PUB-2020-001}.
 
 
\printbibliography

@article{Furry:1937zz,
    author = "Furry, W.H.",
    title = "{A Symmetry Theorem in the Positron Theory}",
    doi = "10.1103/PhysRev.51.125",
    journal = "Phys. Rev.",
    volume = "51",
    pages = "125--129",
    year = "1937"
}

@article{Gabrielli:2007wf,
    author = "Gabrielli, Emidio and Maltoni, Fabio and Mele, Barbara and Moretti, Mauro and Piccinini, Fulvio and Pittau, Roberto",
    title = "{Higgs boson production in association with a photon in vector boson fusion at the LHC}",
    eprint = "hep-ph/0702119",
    archivePrefix = "arXiv",
    reportNumber = "CP3-07-03, FNT-T-2007-01, HIP-2006-01-TH",
    doi = "10.1016/j.nuclphysb.2007.05.010",
    journal = "Nucl. Phys. B",
    volume = "781",
    pages = "64--84",
    year = "2007"
}

@article{Gabrielli:2016mdd,
    author = "Gabrielli, Emidio and Mele, Barbara and Piccinini, Fulvio and Pittau, Roberto",
    title = "{Asking for an extra photon in Higgs production at the LHC and beyond}",
    eprint = "1601.03635",
    archivePrefix = "arXiv",
    primaryClass = "hep-ph",
    doi = "10.1007/JHEP07(2016)003",
    journal = "JHEP",
    volume = "07",
    pages = "003",
    year = "2016"
}

@article{Aad:2020jym,
    author="{ATLAS Collaboration}",
    title = "{Measurements of $WH$ and $ZH$ production in the $H \rightarrow b\bar{b}$ decay channel in $pp$ collisions at 13 TeV with the ATLAS detector}",
    eprint = "2007.02873",
    archivePrefix = "arXiv",
    primaryClass = "hep-ex",
    reportNumber = "CERN-EP-2020-087",
    month = "7",
    year = "2020"
}

@article{LUCID2,
  author={G. Avoni and others},
  title={The new LUCID-2 detector for luminosity measurement and monitoring in ATLAS},
  journal={JINST},
  volume={13},
  number={07},
  pages={P07017},
  doi="10.1088/1748-0221/13/07/P07017",
  year={2018}
}

@Booklet{bukin2007fitting,
      author         = "Bukin, A. D.",
      title          = "{Fitting function for asymmetric peaks}",
      year           = "2007",
      archivePrefix  = "arXiv",
      eprint         = "0711.4449",
      primaryClass   = "physics.data-an",
}

@article{Choudalakis_2012,
   title={Plotting the differences between data and expectation},
   volume={127},
   ISSN={2190-5444},
   url={http://dx.doi.org/10.1140/epjp/i2012-12025-y},
   DOI={10.1140/epjp/i2012-12025-y},
   number={2},
   pages={25},
   journal={Eur. Phys. J. Plus},
   publisher={Springer Science and Business Media LLC},
   author={Choudalakis, G. and Casadei, D.},
   year={2012}
}

@Article{Cacciari:2008gp,
     author    = "Cacciari, Matteo and Salam, Gavin P. and Soyez, Gregory",
     title     = "{The anti-\(k_{t}\) jet clustering algorithm}",
     journal   = "JHEP",
     volume    = "04",
     year      = "2008",
     pages     = "063",
     eprint    = "0802.1189",
     archivePrefix = "arXiv",
     primaryClass  =  "hep-ph",
     doi       = "10.1088/1126-6708/2008/04/063",
     SLACcitation  = "%%CITATION = 0802.1189;%%"
}

@Article{Fastjet,
      author         = "Cacciari, Matteo and Salam, Gavin P. and Soyez, Gregory",
      title          = "{FastJet user manual}",
      journal        = "Eur. Phys. J. C",
      volume         = "72",
      year           = "2012",
      pages          = "1896",
      doi            = "10.1140/epjc/s10052-012-1896-2",
      eprint         = "1111.6097",
      archivePrefix  = "arXiv",
      primaryClass   = "hep-ph",
      reportNumber   = "CERN-PH-TH-2011-297",
      SLACcitation   = "%%CITATION = ARXIV:1111.6097;%%"
}

@Article{Butterworth:2015oua,
      author         = "Butterworth, Jon and others",
      title          = "{PDF4LHC recommendations for LHC Run II}",
      journal        = "J. Phys. G",
      volume         = "43",
      year           = "2016",
      pages          = "023001",
      doi            = "10.1088/0954-3899/43/2/023001",
      eprint         = "1510.03865",
      archivePrefix  = "arXiv",
      primaryClass   = "hep-ph",
      reportNumber   = "OUTP-15-17P, SMU-HEP-15-12, TIF-UNIMI-2015-14,
                        LCTS-2015-27, CERN-PH-TH-2015-249",
      SLACcitation   = "%%CITATION = ARXIV:1510.03865;%%"
}

@article{Lai:2010vv,
      author         = "Lai, H.-L. and others",
      title          = "{New parton distributions for collider physics}",
      journal        = "Phys. Rev. D",
      volume         = "82",
      pages          = "074024",
      doi            = "10.1103/PhysRevD.82.074024",
      year           = "2010",
      eprint         = "1007.2241",
      archivePrefix  = "arXiv",
      primaryClass   = "hep-ph",
      reportNumber   = "MSUHEP-100707, SMU-HEP-10-10",
      SLACcitation   = "%%CITATION = ARXIV:1007.2241;%%",
}

@article{Ball:2012cx,
      author         = "Ball, Richard D. and others",
      title          = "{Parton distributions with LHC data}",
      journal        = "Nucl. Phys. B",
      volume         = "867",
      year           = "2013",
      pages          = "244",
      doi            = "10.1016/j.nuclphysb.2012.10.003",
      eprint         = "1207.1303",
      archivePrefix  = "arXiv",
      primaryClass   = "hep-ph",
      reportNumber   = "EDINBURGH-2012-08, IFUM-FT-997, FR-PHENO-2012-014,
                        RWTH-TTK-12-25, CERN-PH-TH-2012-037, SFB-CPP-12-47\,
                        --CERN-PH-TH-2012-037",
      SLACcitation   = "%%CITATION = ARXIV:1207.1303;%%"
}

@article{Sjostrand:2014zea,
      author         = "Sj{\"o}strand, Torbj{\"o}rn and Ask, Stefan and Christiansen,
                        Jesper R. and Corke, Richard and Desai, Nishita and Ilten,
                        Philip and Mrenna, Stephen and Prestel, Stefan and
                        Rasmussen, Christine O. and Skands, Peter Z.",
      title          = "{An introduction to PYTHIA 8.2}",
      journal        = "Comput. Phys. Commun.",
      volume         = "191",
      year           = "2015",
      pages          = "159",
      doi            = "10.1016/j.cpc.2015.01.024",
      eprint         = "1410.3012",
      archivePrefix  = "arXiv",
      primaryClass   = "hep-ph",
      reportNumber   = "LU-TP-14-36, MCNET-14-22, CERN-PH-TH-2014-190,
                        FERMILAB-PUB-14-316-CD, DESY-14-178, SLAC-PUB-16122,
                        --FERMILAB-PUB-14-316-CD",
      SLACcitation   = "%%CITATION = ARXIV:1410.3012;%%"
}

@Article{Lange:2001uf,
      author         = "Lange, D. J.",
      title          = "{The EvtGen particle decay simulation package}",
      booktitle      = "{Proceedings, 7th International Conference on B physics
                        at hadron machines (BEAUTY 2000)}",
      journal        = "Nucl. Instrum. Meth. A",
      volume         = "462",
      year           = "2001",
      pages          = "152",
      doi            = "10.1016/S0168-9002(01)00089-4",
      SLACcitation   = "%%CITATION = NUIMA,A462,152;%%"
}

@Article{Alwall:2014hca,
      author         = "Alwall, J. and Frederix, R. and Frixione, S. and Hirschi,
                        V. and Maltoni, F. and Mattelaer, O. and Shao, H. -S. and
                        Stelzer, T. and Torrielli, P. and Zaro, M.",
      title          = "{The automated computation of tree-level and
                        next-to-leading order differential cross sections, and
                        their matching to parton shower simulations}",
      journal        = "JHEP",
      volume         = "07",
      year           = "2014",
      pages          = "079",
      doi            = "10.1007/JHEP07(2014)079",
      eprint         = "1405.0301",
      archivePrefix  = "arXiv",
      primaryClass   = "hep-ph",
      reportNumber   = "CERN-PH-TH-2014-064, CP3-14-18, LPN14-066, MCNET-14-09,
                        ZU-TH-14-14",
      SLACcitation   = "%%CITATION = ARXIV:1405.0301;%%"
}

@Article{Bahr:2008pv,
      author         = "B{\"a}hr, M. and others",
      title          = "{Herwig++ physics and manual}",
      journal        = "Eur. Phys. J. C",
      volume         = "58",
      year           = "2008",
      pages          = "639",
      doi            = "10.1140/epjc/s10052-008-0798-9",
      eprint         = "0803.0883",
      archivePrefix  = "arXiv",
      primaryClass   = "hep-ph",
      reportNumber   = "CERN-PH-TH-2008-038, CAVENDISH-HEP-08-03, KA-TP-05-2008,
                        DCPT-08-22, IPPP-08-11, CP3-08-05",
      SLACcitation   = "%%CITATION = ARXIV:0803.0883;%%"
}

@Article{Bellm:2015jjp,
      author         = "Bellm, Johannes and others",
      title          = "{Herwig 7.0/Herwig++ 3.0 release note}",
      journal        = "Eur. Phys. J. C",
      volume         = "76",
      year           = "2016",
      number         = "4",
      pages          = "196",
      doi            = "10.1140/epjc/s10052-016-4018-8",
      eprint         = "1512.01178",
      archivePrefix  = "arXiv",
      primaryClass   = "hep-ph",
      reportNumber   = "CERN-PH-TH-2015-289, MAN-HEP-2015-15, IFJPAN-IV-2015-13,
                        HERWIG-2015-01, KA-TP-18-2015, DCPT-15-142, MCNET-15-28,
                        IPPP-15-71, --HERWIG-2015-01",
      SLACcitation   = "%%CITATION = ARXIV:1512.01178;%%"
}

@Article{Nason:2004rx,
      author         = "Nason, Paolo",
      title          = "{A new method for combining NLO QCD with shower Monte Carlo algorithms}",
      journal        = "JHEP",
      volume         = "11",
      pages          = "040",
      doi            = "10.1088/1126-6708/2004/11/040",
      year           = "2004",
      eprint         = "hep-ph/0409146",
      archivePrefix  = "arXiv",
}

@Article{Frixione:2007vw,
      author         = "Frixione, Stefano and Nason, Paolo and Oleari, Carlo",
      title          = "{Matching NLO QCD computations with parton shower
                        simulations: the POWHEG method}",
      journal        = "JHEP",
      volume         = "11",
      pages          = "070",
      doi            = "10.1088/1126-6708/2007/11/070",
      year           = "2007",
      eprint         = "0709.2092",
      archivePrefix  = "arXiv",
      primaryClass   = "hep-ph",
}

@Article{Alioli:2010xd,
      author         = "Alioli, Simone and Nason, Paolo and Oleari, Carlo and Re,
                        Emanuele",
      title          = "{A general framework for implementing NLO calculations in
                        shower Monte Carlo programs: the POWHEG BOX}",
      journal        = "JHEP",
      volume         = "06",
      pages          = "043",
      doi            = "10.1007/JHEP06(2010)043",
      year           = "2010",
      eprint         = "1002.2581",
      archivePrefix  = "arXiv",
      primaryClass   = "hep-ph",
}

@article{deFlorian:2016spz,
      author         = "de Florian, D. and others",
      title          = "{Handbook of LHC Higgs Cross Sections: 4. Deciphering the
                        Nature of the Higgs Sector}",
      collaboration  = "LHC Higgs Cross Section Working Group",
      doi            = "10.23731/CYRM-2017-002",
      year           = "2016",
      eprint         = "1610.07922",
      archivePrefix  = "arXiv",
      primaryClass   = "hep-ph",
      reportNumber   = "FERMILAB-FN-1025-T, CERN-2017-002-M",
      SLACcitation   = "%%CITATION = ARXIV:1610.07922;%%"
}

@article{Agostinelli:2002hh,
      author         = "Agostinelli, S. and others",
      title          = "{\textsc{Geant4} -- a simulation toolkit}",
      journal        = "Nucl. Instrum. Meth. A",
      volume         = "506",
      year           = "2003",
      pages          = "250",
      doi            = "10.1016/S0168-9002(03)01368-8",
      reportNumber   = "SLAC-PUB-9350, FERMILAB-PUB-03-339",
      SLACcitation   = "%%CITATION = NUIMA,A506,250;%%"
}

@Booklet{TMVA,
      author         = "Hoecker, Andreas and others",
      title          = "{TMVA - Toolkit for Multivariate Data Analysis}",
      year           = "2007",
      eprint         = "physics/0703039",
      archivePrefix  = "arXiv",
      primaryClass   = "physics.data-an",
}

@Article{PERF-2007-01,
    author         = "{ATLAS Collaboration}",
    title          = "{The ATLAS Experiment at the CERN Large Hadron Collider}",
    journal        = "JINST",
    volume         = "3",
    year           = "2008",
    pages          = "S08003",
    doi            = "10.1088/1748-0221/3/08/S08003",
    primaryClass   = "hep-ex",
}

@Article{SOFT-2010-01,
    author         = "{ATLAS Collaboration}",
    title          = "{The ATLAS Simulation Infrastructure}",
    journal        = "Eur. Phys. J. C",
    volume         = "70",
    year           = "2010",
    pages          = "823",
    doi            = "10.1140/epjc/s10052-010-1429-9",
    eprint         = "1005.4568",
    archivePrefix  = "arXiv",
    primaryClass   = "physics.ins-det",
}

@Article{HIGG-2012-27,
    author         = "{ATLAS Collaboration}",
    title          = "{Observation of a new particle in the search for the Standard Model Higgs boson with the ATLAS detector at the LHC}",
    journal        = "Phys. Lett. B",
    volume         = "716",
    year           = "2012",
    pages          = "1",
    doi            = "10.1016/j.physletb.2012.08.020",
    reportNumber   = "CERN-PH-EP-2012-218",
    eprint         = "1207.7214",
    archivePrefix  = "arXiv",
    primaryClass   = "hep-ex",
}

@Article{STDM-2012-23,
    author         = "{ATLAS Collaboration}",
    title          = "{Measurement of the \(Z/\gamma^*\) boson transverse momentum distribution in \(pp\) collisions at \(\sqrt{s} = 7\,\text{TeV}\) with the ATLAS detector}",
    journal        = "JHEP",
    volume         = "09",
    year           = "2014",
    pages          = "145",
    doi            = "10.1007/JHEP09(2014)145",
    reportNumber   = "CERN-PH-EP-2014-075",
    eprint         = "1406.3660",
    archivePrefix  = "arXiv",
    primaryClass   = "hep-ex",
}

@Article{PERF-2014-07,
    author         = "{ATLAS Collaboration}",
    title          = "{Topological cell clustering in the ATLAS calorimeters and its performance in LHC Run~1}",
    journal        = "Eur. Phys. J. C",
    volume         = "77",
    year           = "2017",
    pages          = "490",
    doi            = "10.1140/epjc/s10052-017-5004-5",
    reportNumber   = "CERN-PH-EP-2015-304",
    eprint         = "1603.02934",
    archivePrefix  = "arXiv",
    primaryClass   = "hep-ex",
}

@Article{PERF-2015-10,
    author         = "{ATLAS Collaboration}",
    title          = "{Muon reconstruction performance of the ATLAS detector in proton--proton collision data at \(\sqrt{s} = 13\,\text{TeV}\)}",
    journal        = "Eur. Phys. J. C",
    volume         = "76",
    year           = "2016",
    pages          = "292",
    doi            = "10.1140/epjc/s10052-016-4120-y",
    reportNumber   = "CERN-EP-2016-033",
    eprint         = "1603.05598",
    archivePrefix  = "arXiv",
    primaryClass   = "hep-ex",
}

@Article{HIGG-2016-30,
    author         = "{ATLAS Collaboration}",
    title          = "{Search for Higgs bosons produced via vector-boson fusion and decaying into bottom quark pairs in \(\sqrt{s} = 13\,\text{TeV}\) \(pp\) collisions with the ATLAS detector}",
    journal        = "Phys. Rev. D",
    volume         = "98",
    year           = "2018",
    pages          = "052003",
    doi            = "10.1103/PhysRevD.98.052003",
    reportNumber   = "CERN-EP-2018-140",
    eprint         = "1807.08639",
    archivePrefix  = "arXiv",
    primaryClass   = "hep-ex",
}

@Article{TRIG-2016-01,
    author         = "{ATLAS Collaboration}",
    title          = "{Performance of the ATLAS trigger system in 2015}",
    journal        = "Eur. Phys. J. C",
    volume         = "77",
    year           = "2017",
    pages          = "317",
    doi            = "10.1140/epjc/s10052-017-4852-3",
    reportNumber   = "CERN-EP-2016-241",
    eprint         = "1611.09661",
    archivePrefix  = "arXiv",
    primaryClass   = "hep-ex",
}

@Article{EGAM-2018-01,
    author         = "{ATLAS Collaboration}",
    title          = "{Electron and photon performance measurements with the ATLAS detector using the 2015--2017 LHC proton--proton collision data}",
    journal        = "JINST",
    volume         = "14",
    year           = "2019",
    pages          = "P12006",
    doi            = "10.1088/1748-0221/14/12/P12006",
    reportNumber   = "CERN-EP-2019-145",
    eprint         = "1908.00005",
    archivePrefix  = "arXiv",
    primaryClass   = "hep-ex",
}

@Article{FTAG-2018-01,
    author         = "{ATLAS Collaboration}",
    title          = "{ATLAS \(b\)-jet identification performance and efficiency measurement with \(t\bar{t}\) events in \(pp\) collisions at \(\sqrt{s} = 13\,\text{TeV}\)}",
    journal        = "Eur. Phys. J. C",
    volume         = "79",
    year           = "2019",
    pages          = "970",
    doi            = "10.1140/epjc/s10052-019-7450-8",
    reportNumber   = "CERN-EP-2019-132",
    eprint         = "1907.05120",
    archivePrefix  = "arXiv",
    primaryClass   = "hep-ex",
}

@Article{HIGG-2018-57,
    author         = "{ATLAS Collaboration}",
    title          = "{Combined measurements of Higgs boson production and decay using up to \(80\,\text{fb}^{-1}\) of proton--proton collision data at \(\sqrt{s} = 13\,\text{TeV}\) collected with the ATLAS experiment}",
    journal        = "Phys. Rev. D",
    volume         = "101",
    year           = "2020",
    pages          = "012002",
    doi            = "10.1103/PhysRevD.101.012002",
    reportNumber   = "CERN-EP-2019-097",
    eprint         = "1909.02845",
    archivePrefix  = "arXiv",
    primaryClass   = "hep-ex",
}

@Article{JETM-2018-05,
    author         = "{ATLAS Collaboration}",
    title          = "{Jet energy scale and resolution measured in proton--proton collisions at \(\sqrt{s} = 13\,\text{TeV}\) with the ATLAS detector}",
    year           = "2020",
    reportNumber   = "CERN-EP-2020-083",
    eprint         = "2007.02645",
    archivePrefix  = "arXiv",
    primaryClass   = "hep-ex",
}

@Article{PIX-2018-001,
    author         = "Abbott, B. and others",
    title          = "{Production and integration of the ATLAS Insertable B-Layer}",
    journal        = "JINST",
    volume         = "13",
    year           = "2018",
    pages          = "T05008",
    doi            = "10.1088/1748-0221/13/05/T05008",
    eprint         = "1803.00844",
    archivePrefix  = "arXiv",
    primaryClass   = "physics.ins-det",
}

@Booklet{ATL-SOFT-PUB-2020-001,
    author         = "{ATLAS Collaboration}",
    title          = "{ATLAS Computing Acknowledgements}",
    howpublished   = "{ATL-SOFT-PUB-2020-001}",
    url            = "https://cds.cern.ch/record/2717821",
}

@Report{ATLAS-TDR-2010-19,
    author         = "{ATLAS Collaboration}",
    title          = "{ATLAS Insertable B-Layer Technical Design Report}",
    type           = "ATLAS-TDR-19; CERN-LHCC-2010-013",
    year           = "2010",
    url            = "https://cds.cern.ch/record/1291633",
}

@Article{CMS-HIG-12-028,
    author         = "{CMS Collaboration}",
    title          = "{Observation of a new boson at a mass of 125 GeV with the CMS experiment at the LHC}",
    journal        = "Phys. Lett. B",
    volume         = "716",
    year           = "2012",
    pages          = "30",
    doi            = "10.1016/j.physletb.2012.08.021",
    reportNumber   = "CERN-PH-EP-2012-220",
    eprint         = "1207.7235",
    archivePrefix  = "arXiv",
    primaryClass   = "hep-ex",
}

@Article{CMS-HIG-14-004,
    author         = "{CMS Collaboration}",
    title          = "{Search for the standard model Higgs boson produced through vector boson fusion and decaying to \(b \bar{b}\)}",
    journal        = "Phys. Rev. D",
    volume         = "92",
    year           = "2015",
    pages          = "032008",
    doi            = "10.1103/PhysRevD.92.032008",
    reportNumber   = "CERN-PH-EP-2015-121",
    eprint         = "1506.01010",
    archivePrefix  = "arXiv",
    primaryClass   = "hep-ex",
}

@Article{CMS-HIG-17-031,
    author         = "{CMS Collaboration}",
    title          = "{Combined measurements of Higgs boson couplings in proton--proton collisions at \(\sqrt{s} = 13\,\text{TeV}\)}",
    journal        = "Eur. Phys. J. C",
    volume         = "79",
    year           = "2019",
    pages          = "421",
    doi            = "10.1140/epjc/s10052-019-6909-y",
    reportNumber   = "CERN-EP-2018-263",
    eprint         = "1809.10733",
    archivePrefix  = "arXiv",
    primaryClass   = "hep-ex",
}

@Article{CMS-HIG-18-016,
    author         = "{CMS Collaboration}",
    title          = "{Observation of Higgs Boson Decay to Bottom Quarks}",
    journal        = "Phys. Rev. Lett.",
    volume         = "121",
    year           = "2018",
    pages          = "121801",
    doi            = "10.1103/PhysRevLett.121.121801",
    reportNumber   = "CERN-EP-2018-223",
    eprint         = "1808.08242",
    archivePrefix  = "arXiv",
    primaryClass   = "hep-ex",
}

@Booklet{ATLAS-CONF-2014-018,
    author         = "{ATLAS Collaboration}",
    title          = "{Tagging and suppression of pileup jets with the ATLAS detector}",
    howpublished   = "{ATLAS-CONF-2014-018}",
    url            = "https://cds.cern.ch/record/1700870",
    year           = "2014",
}

@Booklet{ATLAS-CONF-2018-001,
    author         = "{ATLAS Collaboration}",
    title          = "{Measurement of \(b\)-tagging efficiency of \(c\)-jets in \(t\bar{t}\) events using a likelihood approach with the ATLAS detector}",
    howpublished   = "{ATLAS-CONF-2018-001}",
    url            = "https://cds.cern.ch/record/2306649",
    year           = "2018",
}

@Booklet{ATLAS-CONF-2018-006,
    author         = "{ATLAS Collaboration}",
    title          = "{Calibration of light-flavour \(b\)-jet mistagging rates using ATLAS proton--proton collision data at \(\sqrt{s} = 13~\text{TeV}\)}",
    howpublished   = "{ATLAS-CONF-2018-006}",
    url            = "https://cds.cern.ch/record/2314418",
    year           = "2018",
}

@Booklet{ATLAS-CONF-2019-021,
    author         = "{ATLAS Collaboration}",
    title          = "{Luminosity determination in \(pp\) collisions at \(\sqrt{s} = 13\,\text{TeV}\) using the ATLAS detector at the LHC}",
    howpublished   = "{ATLAS-CONF-2019-021}",
    url            = "https://cds.cern.ch/record/2677054",
    year           = "2019",
}

@Booklet{ATL-PHYS-PUB-2015-022,
    author         = "{ATLAS Collaboration}",
    title          = "{Expected performance of the ATLAS \(b\)-tagging algorithms in Run-2}",
    howpublished   = "{ATL-PHYS-PUB-2015-022}",
    url            = "https://cds.cern.ch/record/2037697",
    year           = "2015",
}

@Booklet{ATL-PHYS-PUB-2016-017,
    author         = "{ATLAS Collaboration}",
    title          = "{The Pythia~8 A3 tune description of ATLAS minimum bias and inelastic measurements incorporating the Donnachie--Landshoff diffractive model}",
    howpublished   = "{ATL-PHYS-PUB-2016-017}",
    url            = "https://cds.cern.ch/record/2206965",
    year           = "2016",
}
 
\clearpage
 
\clearpage
 
\begin{flushleft}
\hypersetup{urlcolor=black}
{\Large The ATLAS Collaboration}

\bigskip

\AtlasOrcid[0000-0002-6665-4934]{G.~Aad}$^\textrm{\scriptsize 101}$,    
\AtlasOrcid[0000-0002-5888-2734]{B.~Abbott}$^\textrm{\scriptsize 127}$,    
\AtlasOrcid[0000-0002-7248-3203]{D.C.~Abbott}$^\textrm{\scriptsize 102}$,    
\AtlasOrcid[0000-0002-2788-3822]{A.~Abed~Abud}$^\textrm{\scriptsize 36}$,    
\AtlasOrcid[0000-0002-1002-1652]{K.~Abeling}$^\textrm{\scriptsize 53}$,    
\AtlasOrcid[0000-0002-2987-4006]{D.K.~Abhayasinghe}$^\textrm{\scriptsize 93}$,    
\AtlasOrcid[0000-0002-8496-9294]{S.H.~Abidi}$^\textrm{\scriptsize 29}$,    
\AtlasOrcid[0000-0002-8279-9324]{O.S.~AbouZeid}$^\textrm{\scriptsize 40}$,    
\AtlasOrcid{N.L.~Abraham}$^\textrm{\scriptsize 155}$,    
\AtlasOrcid[0000-0001-5329-6640]{H.~Abramowicz}$^\textrm{\scriptsize 160}$
% ...
\bigskip
\\

% ...
$^{29}$Physics Department, Brookhaven National Laboratory, Upton NY; United States of America.\\
% ...
$^{36}$CERN, Geneva; Switzerland.\\
% ...
$^{40}$Niels Bohr Institute, University of Copenhagen, Copenhagen; Denmark.\\
% ...
$^{53}$II. Physikalisches Institut, Georg-August-Universit\"{a}t G\"ottingen, G\"ottingen; Germany.\\
% ...
$^{93}$Department of Physics, Royal Holloway University of London, Egham; United Kingdom.\\
% ...
$^{101}$CPPM, Aix-Marseille Universit\'e, CNRS/IN2P3, Marseille; France.\\
$^{102}$Department of Physics, University of Massachusetts, Amherst MA; United States of America.\\
% ...
$^{127}$Homer L. Dodge Department of Physics and Astronomy, University of Oklahoma, Norman OK; United States of America.\\
% ...
$^{155}$Department of Physics and Astronomy, University of Sussex, Brighton; United Kingdom.\\
% ...
$^{160}$Raymond and Beverly Sackler School of Physics and Astronomy, Tel Aviv University, Tel Aviv; Israel.\\
% ...

\end{flushleft}

 
\end{document}